A New Disintegrative Capture Theory for the Origin of the Moon

Peter D. Noerdlinger[1]
1222 Oakleaf Circle
Boulder, CO 80304-1151
April 8, 2012

ABSTRACT

Prior to the discovery that the Moon has an Oxygen isotope ratio very close to that of the Earth and very small iron core, a number of researchers (Gerstenkorn 1955, 1969, Singer 1968, Öpik 1972, Mitler 1975) worked out details of a possible capture process. Both the similarity of the Oxygen isotope ratio $^{18}O/^{16}O$ to that of Earth and the smallness of the actual lunar iron core led to virtual abandonment of whole-body capture theories, though some work continued (Conway 1982, Wood 1986, Malcuit et al 1989). The Oxygen problem was that the pre-capture Moon would have necessarily been formed in a region close to 1 A.U. from the Sun, which seemed restrictive, in that the formation of the Earth was expected to eat up or drive away material near its orbit. Further, there is no obvious way to get rid of an original lunar iron core ~32% by mass. The present lunar core is < 4% of its mass. Here, I will show how Moon arose from a proto-Moon that started at L4 and was captured into a descending orbit. Connors et al (2011) actually recently discovered an Earth-Trojan object, of diameter "several hundred meters," named TK$_7$, at L4. The abandoned disintegrative capture theories (Öpik, Mitler, Wood) all involved disintegration *during* capture. Our disintegrative theory puts the disintegration *after* the capture. Capture is into an eccentric orbit, but tidal forces circularize the orbit. The theory was in many ways anticipated by Singer and by Öpik. The location of the proto-Moon before capture has been identified by Belbruno & Gott (2005, hereinafter BG) as either the L4 or L5 Lagrange point of Earth. These authors, however, ignored tidal forces and calculated escape of a Mars-mass planet "Theia" from (e.g.) L4, followed by *collision* with the Earth, as required in the "Giant Impact" (GI) theory (Hartmann & Davis 1975, Cameron & Ward 1976, Canup 2004a,b). Our PM has mass ~1/39 that of Earth or about four times less than Theia. Capture occurs inside the co-rotation radius, so that, just as Mars' moon Phobos approaches its home planet, the PM descends towards Earth. The outer layers of the PM are stripped to form a disk of rock particles. All the rock is in the disk by the time the PM remnant reaches 2.234 Earth radii ($R_E$), after which the iron core disintegrates into an iron ring. The core, having driven much of the rock disk outside the Roche limit, tidally coaxes the disk to form into a single Moon (our present Moon) at about 3.6 - 3.8 Earth radii. The latter then tidally drives the core, which disintegrates into an iron ring, and any portion of the stripped rock that had lain within the orbit of the PM as it lost mass down to the surface of the Earth, possibly producing the "Late Veneer" (O'Neill 1991). The inner iron ring and/or inner rock disk particles arrive at the Earth's equator (of date) at ~ 8 km s$^{-1}$ and virtually zero angle of incidence. This contrasts to the GI case, where Theia's core enters at near-normal incidence and descends rapidly to join the Earth's core (thus having no discernable contribution to the Late Veneer.) In the present theory, as contrasted to the GI case, the lunar material is never vaporized. Thus, magnetized rocks can survive from the PM state, where the iron core was big enough to form a dynamo (Glatzmeier, personal communication, 2011). The age of the magnetized Troctolite 76535 at 4.291 Ga therefore predates the capture event. The recent determination from the magnetization and age of basalt 10020 (Shea et al 2012) that the lunar dynamo persisted to ~ 3.72 Ga B.P. causes one to consider that date also as a time frame for the capture. The last parcels of rock to tear free from the PM's core might be driven out and hit the Moon on the nearside, so as to form the maria at the same time, within ~ a few days, as the PM core plasters the Earth's surface with the veneer. This would be the "Late Veneer" only in Case III (see below). The theory requires capture, and, in honor of the early work of Horst Gerstenkorn the capture event will be denoted the "GE" – or Gerstenkorn Event. With this event - the formation of the Moon as we know it at possibly at only 3.72 - 3.8 Ga ago, the classic problem of too-rapid outward evolution of the orbit, leading to the Moon's having been very near Earth less than 2 Ga ago (Kaula 1968) is ameliorated. That discrepancy has prompted attempts to model the ocean basins in the Pre-Cambrian period and extensive analyses of geologic evidence (Williams 2000). Furthermore, the origin of life in the Archaen epoch and not within the Hadean jibes with the Moon's having formed near their boundary ~3.8 Ga ago (Ryder, 2002). The theory presented here is compatible with lunar ages of 4.6 Ga plus (Case I), 4.291 Ga (Case II, wherein the magnetization of Troctolite 67535 is explained) or 3.8 Ga or a bit less (Case III) wherein the magnetizations of both that rock and lava 10020 are explained. No matter which date proves to be right, the Moon was essentially turned inside-out by the processes discussed here, because the outermost layers were forced out of the Roche zone first and began to collect into the actual Moon, while the inner layers of the PM were driven out of the Roche zone or limit later, finally accreting to form today's Moon. The restructuring process had a random component due to circulation or turbulence in the disk outside ~3.6 Earth radii where the Moon re-accreted. This explains many cases of the juxtaposition of rocks or lavas of different ages as found in the Apollo program.

---

[1] Consultant, Technology Advancements, Inc. Playa del Rey, CA

The prevalent theory for the formation of the Moon is the Giant Impact (GI) theory (Hartmann & Davis 1975, Cameron & Ward 1976, Canup & Asphaug 2001, Boss et al 1991). The GI theory deals with the so-called "iron problem" in that the Moon has only a small iron (or other - see Wieczorek & Zuber 2002) core, ~2% - 4% of its mass. When combined with the theory of Belbruno & Gott (2005) the GI theory also solves the Oxygen Isotope problem, namely that the Earth and Moon have similar Oxygen isotope ratios (Clayton, 1973), while objects formed closer to or farther from the Sun differ in this ratio. Nevertheless, the lunar mantle is ~ 50% - 60% richer in iron than the Earth's

(Jones & Palme, 2000), a fact not explained in the GI theory. It is possible to suppose that the Moon was formed from Earth mantle material richer than now in iron because Earth's core might not have formed, although Canup (2004b) gives the age of core formation as very early, as do Yin et al (2002), and Edmunson et al (2009). Simulations supporting the GI theory, however Canup (2004), show that the Moon formed from impactor material, not Earth mantle material, although, in another contradiction, (Jones and Palme, 2000) aver, based on the Hf/W ratio being the same as Earth's and the $2\varepsilon$ W isotope anomaly in KREEP-rich basalts, that they "see no way to make the Moon solely from terrestrial mantle material." Similarly, Zindler & Jacobsen (2010) object to the melting of Earth's mantle, numerous problems with isotopes and to the assumed eddy size in Pahlevan and Stevenson (2007). While the theory presented here does not offer a well-defined way to seed the lunar mantle with 50% - 60% more iron by mass than Earth's, there are two possibilities. One would be that the PM's core formation was incomplete, leaving more iron in its mantle. The other would be that towards the end of rock stripping from the PM, a certain amount of core material was torn off with the rock and mixed in. This is particularly reasonable in that the core would be much deformed by tidal forces. The first explanation runs counter to Hf-W evidence that the PM's core was formed early.

It is surprising that various lines of approach on the study of our permanent Earth satellite by capture (Gerstenkorn 1955, 1969, Singer 1968, Öpik 1972 (hereinafter O72), Mitler 1975, Conway 1982, Malcuit et al 1989), by collision (GI theory), and a third line of research on the existence of co-orbital "temporary" satellites (Wiegert at al 1997, Namouni et al 1999, Connors et al 2002, Connors et al 2004, 2011, Brasser et al 2004, Mikkola et al 2006, Morais & Morbidelli 2002, Wajer 2009, Christou and Asher 2011) as well as on the navigation of spacecraft (Belbruno 2004) have had limited intersection. The first group of authors showed that, dynamically speaking, our present Moon could have been captured into orbit. The second group avoided capture theories as seemingly unworkable, due to the iron core and Oxygen isotope problems, and simply required the "impactor," of mass about that of Mars, to directly strike the Earth. The third group had no reason to consider tidal forces, which would be extremely small for asteroids of mass $< 10^{10}$ kg at a distance of 0.01 to 0.1 AU, as is common for our "co-orbital" or "visiting" satellites, as well as for spacecraft. There is a lot of "grey area" between the masses of tiny asteroids and that of Mars, however, and tidal forces tend to dominate in some cases, depending on impact parameter, for masses equal to or larger than our present Moon. Otherwise, Gerstenkorn, Conway and others would not have found capture. Conway's equations involve averaging change rates of orbital elements over true anomaly, which does not constitute averaging over time. We have therefore done direct numerical integrations of the orbits rather than using his averages. It can be argued that previous authors had to assume contrived initial conditions in order to effect capture. We have two ways out of that problem. For one thing, our initial body is 70% to 100% more massive than the Moon (it will lose ~9% - 20% of its original mass in becoming the Moon). Tidal forces are greater on more massive bodies. The second "escape hatch" from being challenged as to "contrived" initial conditions is that the studies of co-orbital satellites show that one can call on repeated encounters. In other words, the original capture theories tended to discount problems with the orbit of the PM after one dissipative pass still reaching outside the Hill radius, ~ 0.01 AU. We now see from co-orbital satellite theory that several small asteroids remain within ~0.1 AU of the Earth for numerous orbits, before returning to horseshoe orbits around the Sun. For low-mass asteroids this is of interest, but has no decisive effect. For a proto-Moon (PM), it is important, because such an object will repeatedly be subject to tidal dissipation, lowering the apogee. Thus, the allowable range of initial (hyperbolic) impact parameters for capture is extended. Belbruno & Gott (2005) also found repeated Earth encounters. Schmitt (2006) has explicitly called for a new lunar capture and evolution theory, based on his observations as the only geologist to have set foot on the Moon, and his analysis of lunar geology. Schmitt, who prefers an intact-capture theory, asserts that the Moon has a transition to a lower mantle of unconsolidated rock at ~550 km depth, inconsistent with formation from a cloud of vapor (also see Watters et al 2012). Unconsolidated rock seems to this author a more likely source of deep moonquakes (Goins et al 1977) (Lognonne 2005) due to slip under tidal stresses or simple relaxation than rock condensed *en masse* out of vapor or remelted after solidification. Schmitt (2002) states that lower mantle also has volatile element concentrations 10 to 100 times greater than in the upper mantle and crust, again in disagreement with the GI concept. Schmitt (2006) states that isotopes of rare earth elements and of potassium are not fractionated, as one would expect due to fractional crystallization at the high temperatures present in the GI theory. Schmitt skirts the problem of missing core iron in the Moon by suggesting that before its capture it was formed by collision of two bodies, with the iron cores knocked out. This leaves a problem with the oxygen isotope ratios matching Earth, because Schmitt's colliders seem to have been formed further from the Sun. Finally, Schmitt points out that the lower lunar mantle is significantly more aluminous than the upper, and he points out that a generation of dark-haloed cryptomaria (Bell & Hawke 1984) establish a more extended bombardment than usually assumed in the GI theory. Lee et al (2007) also find, based on $^{182}$Hf-$^{182}$W analysis of 21 samples, that the Moon's mantle is poorly mixed, consistent with unconsolidated rock and not, in this author's opinion, with condensation of rock out of vapor. Water has been found in lunar glass beads at an initial 745 ppm level by Saal et al (2008) and at ~1,600 ppm in an apatite rock by Boyce et al (2010). This appears to be inconsistent with the high temperatures generated in the GI theory. Sharp et al (2010) show, however, that highly variable Cl isotope ratios imply that the lunar mantle is anhydrous with the exception of a few anomalies such as the aforementioned beads. From our standpoint, the situation with water abundance is one more indication of heterogeneity in lunar minerals, which is to be expected in a scenario where rock was stripped from the PM in fragments, virtually scrambled up in a disk, and re-accreted outside the Roche limit. There are recent results that show that Oxygen and Titanium isotope ratios in Moon rocks match so closely to terrestrial values leading to claims that the Moon virtually had to be derived from Earth matter. This problem has been addressed within the GI theory by Pahlevan & Stevenson (2007) (PS), who invoke mixing of impactor and Earth material in the very hot post-impact disk. PS did consider the BG proposal that Theia's isotopic ratios might match Earth's because it was also formed at 1 A.U. from the Sun, but they rejected it based on theories of planetary accretion and a discussion of Mars. Most of the discussion of isotope ratios is based on the GI theory, which tends to form the Moon out of impactor material. Despite the PS reasoning, it is not clear to this author why an impactor formed at L4, i.e. at 1 A.U. from the Sun, should have isotopic composition different from Earth's. Furthermore, though the laboratory work is accurate, the sample of Moon rocks (including, for example, Allan Hills meteorites) offers a limited selection that may not perfectly represent bulk lunar isotope ratios. Kramer et al (2008) found $TiO_2$ abundances 1.5 wt% to 5 wt% in different lunar samples. Also they find that the lunar mantle is inhomogeneous. Saal (personal communication, 2011) has noted that the Ti variability "is produced during the differentiation of the Moon magma ocean after formation it is not an original characteristic of the accretion," which again assumes the GI theory. Granting this interpretation, following Kramer et al, and Giguere et al (2000), it is still hard to validate that the gross lunar Ti isotopic signature is identical to Earth's. There may also have been no magma ocean in the Moon (Borg et al 2011). In conclusion, there being no objects besides the Earth and the PM (either the BG one or ours) known to have formed at 1 A.U. from the Sun, it would not seem surprising if the lunar and terrestrial isotopic ratios matched closely, other than that in the BG theory there is some loss of volatiles that could be mass-specific when diffusion is involved.

A word of caution on tidal forces: The dominant tidal forces are (Gerstenkorn 1955, Conway 1982) a radial force due to deformation of and dissipation in the PM, a radial force due to deformation of the Earth, and a tangential force due to deformation of the Earth (the ordinary "tidal force" now driving the Moon outwards.) The forces due to deformation of the Earth are proportional to the *square* of the orbiter's mass, so that per unit orbiter mass they are *linearly* proportional to the mass. Thus a Mars mass PM is subject to 8 times the tangential tidal accelerations that the Moon would be. The radial tidal force due to dissipation in the satellite is only linearly proportional to its mass (G69) and tends to dominate the tangential tidal force for a PM mass equal to the Moon's (today). The linear proportionality to the satellite's mass, while formally true, is overpowered in practice by a proportionality to the fifth power of its radius, $R_{PM}^5$, so that for similar densities the larger satellite is more affected than the smaller. Boss et al (1991) were concerned with tidal forces on a passing planetesimal disrupting it, but ignored the effect of tidal forces on the orbit, probably because they dealt with a very small mass (0.1 $M_\oplus$). I will show that for a PM of mass ~ $M_\oplus$/39 the tidal forces do the job of capture for a reasonable initial orbit, as was already shown by Gerstenkorn. Boss & Peale (1986) actually state that tidal disruption would not occur on a single close pass, which is fine with the theory presented here, but they, like all other authors, do not allow for whole capture followed by tidal stripping. O72 came close to our theory in the sense that he envisioned rings stripped from the incident body (PM) forming into the Moon, but he did not allow for a partially stripped PM orbiting inside the rings.

It is worth comparing the various approaches to lunar origin in terms of "backwards" vs "forwards" calculations. Goldreich (1966) and Mignard (1979, 1980) attempted to work the problem by tracing the present orbit *backwards* in time, as did, to some extent, Gerstenkorn (1955). This effort is doomed to failure, as uncertainties in the lunar and terrestrial structure, poorly known perturbations[2] (by other planets and asteroids, and collisional accretion to the Moon or Earth) lead to large uncertainties in the primordial

---

[2] The uncertainties in planetary perturbations are due to the radial migration that is believed to have altered many of their major axes

orbit. Other authors, such as Singer, BG, and Malcuit et al began with plausible initial conditions and evolved the system forwards, which is likewise our approach.

Recently, Belbruno & Gott (2005) have shown how the problem of forming the PM at ~1 AU from the Sun, can be solved by forming the PM at the Earth's L4 or L5 point. This would avoid any problem with Oxygen isotopes. Additionally, the author has presented a schema (Noerdlinger 2007) wherein the Moon co-formed with Earth in a process along the lines of that worked out by Harris & Kaula (1975) and had its iron core ripped out in a stripping process due to approach within the Roche limit, because its original orbital period was slightly less than a sol. All of the author's attempts to modify the Harris-Kaula scenario to produce a more massive proto-Moon (hereafter "PM") with a ~32% or more by mass iron core were frustrated, however, by the shortness of the time period for formation, violating mineralogical and isotopic constraints, and by the tendency of the PM either to move far outside the Roche limit or fall into the Earth before accretion could be completed. Harris (1978) has, however, presented a co-accretion model that keeps the iron content small. We remark here that the proto-lunar mass required for the Noerdlinger (2007) scenario would have been neither ~1/81 Earth mass $M_\oplus/81$ as the Moon is today, nor 1/10 Earth mass as the impactor was supposed to be, but rather ~1/39 Earth mass, which is the present lunar mass complemented by an initial lunar iron core, later to be removed, and some extra rock mantle that will accrete to Earth. There the problem stood until the work of BG appeared. These authors constructed an impactor of mass similar to that of Mars, as required by the widely accepted GI theory. In that theory, most of the Moon's mass presumably comes from Earth mantle material, while the impactor's iron core buries itself in the Earth, descending to the latter's core. The impactor's mantle may mix with the Earth's to form the Moon by condensation out of vapor. Let us call the combined Belbruno, Gott, Hartmann, Davis, Cameron, Ward, Canup theory the GI theory, while the L4 formation theory for the impactor will be the BG theory or scenario. That theory satisfies the more recent GI requirement (Canup 2004) of collision at essentially parabolic velocity by calculating a "breakout" from the L4 point followed by tadpole orbits until chaotic variations cause the required collision. When no confusion results PM will stand for the lunar precursor – of mass equal to the Moon's (Gerstenkorn, Conway) or 70% to 100% larger (this work), or $> \approx$ that of Mars (Canup 2004, BG), though the latter may be called "Theia" for clarity when needed. BG obtained breakout by allowing considerable mass growth (to Theia mass) but it seems to us that perturbations due to planetary migration (Gomes et al 2005), though gradual, might have effected breakout at a lower mass.

Finally, there is another hypothesis for lunar origin by de Meijer and van Westrenen (2009) and van Westrenen et al (2012), requiring a nuclear reaction in the deep Earth.

## Age of the Proto-Moon

As in BG, the PM can be assumed to have formed early at the L4 point. Its age can be surmised from the age of the oldest Lunar rocks (Borg et al 2011, Touboul et al 2009) as 50-100 Ma after CAI formation, or about 4.55 - 4.56 Ga ago. There are few constraints within our theory on this age. The oldest Moon rocks or soils, in terms of radioactive dating or radioisotope "closure" can be assumed to date the PM (*not* the Moon). The date of the GE event must be that of the oldest substantial *structures* on the Moon, give or take a few hundred years. That is because the GE event destroyed most of the structure of the PM, leaving only rocks of such size that could resist the process of disintegration of the crust and mantle under tidal forces. Although is if often stated that the age of the lunar surface is that just quoted for the PM, Jutzi and Asphaug (2011), for example, claim that the whole farside could have been created by an accretive impact 200 Ma later, which would be timewise nearer to our proposed GE event (Case I). Schmitt (2006) states that the Procellarum crater is 4.3 Ga old and South Pole-Aitken almost as old. These ages are based on crater counts or damage to the Procellarum rim, but Greeley et al (1993) put Procellarum at 3.18 Ga and South Pole-Aitken at 3.64 Ga, based on cratering. Schmitt also cites zircon crystallization ages 4.3 Ga for some Procellarum "KREEP-related" samples that are "possibly related to the Procellarum event." Ages of structures considerably older than lunar basalt 10020 at $3.72 \pm 0.04$ Ga (Shea et al 2012), which replaces Troctolite 76535 at 4.291 Ga as the last sample of a lunar core magnetic field can pose a constraint or problem for the theory presented here, but the assertions of extreme ages such as 4.3 Ga based on cratering are obviously open to question, and mineral crystallization ages do not necessarily define the age of a *structure*. Recalling the race in the 1960's to 1980's to set the smallest possible value for the Hubble constant and so the oldest age for the Universe, as promulgated in various papers by G. Tammann and A. Sandage (see Bhathal 2012), countered by a more balanced approach, we might be more cautious about derivations of extreme ages than of more widely-based ages (Schmidt 1993). Stöffler et al (2006) give an age for Procellarum of only 3.15 to 3.18 Ga, and, importantly said that they were "discarding" the age $4.35 \pm 0.1$ Ga for the "ancient highlands" originally presented by one of the very authors (Ivanov) and similar ages for "the most densely cratered province" and "the uplands," which are supported by "no firm geologic evidence" and no "clear basis" in isotope data. The appearance of the work of Shea et al after the present research and paper were essentially completed has, on the one hand, put stress on the theory, by tempting us to stretch certain error bars to $2\sigma$ or perhaps $3\sigma$, but on the other hand it squeezes the time frame of our capture process to be almost precisely 3.86 Ga B.P., the age of the Nectaris basin. Ryder (1990) has pointed out that there are no established impact melts older than 3.85 Ga. There are many individual rocks and lavas known to exhibit radioisotope ages ~ 4.0 - 4.4 Ga (Edmunson et al 2009, Meyer 2010). For the theory presented here to work optimally, either a new dynamo had to form after the troctolite age 4.291 Ga (see next section) or, as we prefer, only the ages of the *minerals* are represented ("closure" or crystallization/recrystallization ages) in the time frame before 3.85 Ga, but *structures* (say in the km size range) are not. I claim that the older ages represent events in the PM, not in the Moon as we know it. The worst challenges to a recent date for the GE (formation of the Moon) appear to be the 48 km Descartes Formation and Norite 78238, which was part of the Station 8 Boulder (Meyer 2010, Edmunson et al 2005). Stöffler et al (2006) date Descartes at 3.87 - 3.89 Ga, while Stöffler et al (1985) had found ages up to almost 4.1 Ga. The Station 8 "Boulder" was originally only 0.5 m in diameter, coated in brown glass having a low Co/Ni ratio (which was suggested to imply "meteoritic contamination") and had evidence of shock events. There were *two* (sic) crystallization ages, the original at $4.426 \pm 0.065$ Ga and then a shock "disturbance" at $3.93 \pm 0.21$ Ga of about 300 - 400 kbar (30 - 40 Gpa) pressure.

Going only 1σ on the young side gives us 3.72 Ga, which agrees with the (Shea et al 2012) age! So it is our contention that the Station 8 Boulder crystallized at ~4.426 Ga in the PM, survived the tidal stripping, possibly being impacted by "meteoritic" material, which was actually just other rocks stripped from different layers of the PM, and re-accreted to the Moon itself at ~ 3.72 Ga. The fact that Station 8 Boulder was collected at the surface and had lain at a shallow depth before being excavated on the Baby Ray event, though it had solidified at "plutonic" depth agrees with our concept (see below) that the Moon was largely turned inside out at the GE. The glazing may have been due to proximity to the PM core, meaning that this boulder was one of the last pieces of rock stripped. The shock cracking can be used to determining a very approximate collision velocity. Housen et al (1991) found that a collision at velocity ~45 ms$^{-1}$ (converted from their erg gm$^{-1}$) would catastrophically disrupt an asteroid. Since our boulder was badly shocked but not disrupted, we expect a collision at perhaps ~ 20 ms$^{-1}$ occurred. We do not know if this happened just after stripping or just before final accretion. Tantalizingly, our *Mathematica* model for stripping shows the PM orbit moving in at ~ 1700 ms$^{-1}$ towards the end of rock stripping (a ~ 2.25 $R_E$) with a velocity difference from the previous orbit of ~ 22 ms$^{-1}$. Of course, the collision could not have led to sticking for long, both because of bounce and Roche tidal force. Baldwin (1971) is widely quoted as specifying ages of lunar craters up to 4.5 Ga, but in a footnote to his Table II he states that this number is arbitrary. In conclusion, it is our suspicion that lunar scientists have tended to assign the age of a lunar structure to the age(s) of rocks or lavas in it, sometimes with puzzlement over discordant ages, while in our view it is likely that the discordances are due to the mixing-up of older components that crystallized in the PM with newer ones that date from shock or remelting at the GE event. Readers who distrust this view can assume that the GE event was at ~4.291 or even 4.5 Ga ago and that the magnetized rocks (see next Section) acquired their state from lunar dynamos driven by something other than convection in an iron-nickel (-sulfur) core.

## Magnetized rocks and the Age of the Moon (timing of the GE event)

Garrick-Bethell et al (2009) find evidence for an early lunar magnetic core. Presumably it is hard to make a dynamo with 4% of the satellite's mass so this would tend to support my claim that the primitive Moon or Proto-Moon (hereinafter "PM") had an iron core ~32% by mass. For a different view see Dwyer et al (2011). Lee et al (2007) state that troctolite 76535, which is magnetized, was formed within the first 30 Ma of the solar system. Our theory would ascribe the origin of this rock's magnetism to a dynamo within the PM, which had a 32% iron core. The rock would have had to survive the GE event. On the other hand, Hood (2011) finds aligned remanent magnetism in four impact basins, which contradicts the supposition that the magnetism was generated before the GE event, because we suppose the rocks to have been mixed later. It is possible, however, that this particular set of magnetic impact melt rocks was generated after the GE event, the magnetism being due to the magnetic field on the then-nearby Earth. The present Earth field would be within an order of magnitude of what would have been required to magnetize these rocks at an Earth-Moon distance of 3 to 4 $R_e$. Additionally, Touboul et al (2007) assert on the basis of the $^{182}W/^{184}W$ ratio in two (sic) samples, that the Moon and

Earth formed $62^{+90}_{-10}$ My after the formation of the Solar System. This contrasts with the assertion by Lee et al (2007) that their sample possibly formed "within the first 30My of the solar system," and more compellingly, that their sample of 21 lunar rocks exhibits "tungsten isotopic heterogeneity," as detailed in their footnote 26, implying a range of ages as shown in their figures 1 and 3. I would claim that there should be a range of ages from the time of formation of the PM (perhaps a few tens of millions of years after the solar system formed) to the date of the GE event, which would be the 62 Ma value of Touboul et al. The picture is then that the oldest rocks formed within 30 Ma of the Solar System's origin, while the GE event probably dates to the Late Veneer on Earth 3.9 Ga to 3.8 Ga ago (Holzheid et al 2000; Ryder 2002) but see Frost et al (2004). Early lunar magnetism is discussed by Stock and Woolfson (1983) who developed several theories, including dynamo action, while their Fig. (1) gives a history from rock analyses showing $\sim 10^{-4}$ T at 4 Ga in the past, decaying to a few $\times 10^{-6}$ T now. Their Fig. (1) seems to be not inconsistent with a sudden drop in the magnetic field at the GE event. Glatzmeier (2011) confirms that it is possible to have formed a dynamo like Earth's or Mercury's in a PM of mass $\sim 1/40$ of the Earth's. Our favored mass for the PM is almost half that of Mercury! Le Bars et al (2011) have a dynamo theory based on impacts generating fluid flows in the (present, very small) core, leading to dynamo action. Stegman et al (2003) have an alternate theory for a dynamo driven by mantle convection, while Dwyer et al (2011) offer a theory based on mechanical stirring due to Earth tides. Shea et al claim, however, that Stegman et al's dynamo may not last long enough, while Dwyer et al's as well as Le Bars et al's fields are too weak. Yet another view for the origin of magnetic anomalies is presented by Wieczorek et al (2012). Garrick-Bethell et al (2009) find that Troctolite 76535 was exposed to a $10^{-4}$ T magnetic field $\sim 4.2$ Ga ago. The age of this rock is 4.291 Ga, and its cosmic ray exposure age is $\sim 200$ Ma. (Meyer 2003). Lunar basalt 10020 at $\sim 3.72$ Ga (Shea et al 2012) now replaces the troctolite as the oldest Moon rock with evidence for exposure to a "stable" magnetic field $> \sim 1.2 \times 10^{-5}$ T. In view of the stress that so late a dynamo places on our theory, let us first extend their age on the "old" side by $2\sigma$ to 3.80 Ga. Our date for removal of the PM's core by catastrophic tidal stripping is then $\sim 3.80$ Ga and is before the oldest established structures as per Stöffler et al (2006) and Ryder (1990). Alternatively, we can have a solution less constrained timewise if take one of the other theories for the late (Shea et al) magnetic field; in that case our GE could have happened at 4.291 Ga. Because of the advantages of ameliorating the orbit time problem of the Moon's seemingly have been too close $\sim 2$ Ga ago, and of having a tightly constrained theory, our tendency is to accept Shea et al's date. If the 3.8 Ga age for the Moon is contradicted by some proof that substantial lunar structures (such as the Cordillera or Rook rings) are older, then the fallback position is that 67535 dates the end of the original dynamo and a later one affecting basalt 10020 was produced by the mechanisms of Dwyer et al, Stegman et al or LeBars.

**Interim Storage of the PM**

If we choose Case I or Case II there is little problem in assuming that the PM simply traveled in horseshoe or even breakout orbits (full circle round the Sun) for a few hundred million years (Mikkola et al 2006). In case perturbations (e.g. by Venus) disturb the orbit, the PM might enter the L5 region, remain for a while, and even go back to L4.

In Case III we need to find a way for the PM to hang around for 760 Ma. Wajer (2010) found that one quasi-satellite of Earth could be stable for $10^4$ years, while Christou & Asher (2011) found stability of another horseshoe companion to $10^5$ y. All these times are admittedly short of what we need for Case III, but we point out some more possible scenarios: (a) our PM might have been one out of hundreds of objects ("clones") all but one of which were lost. (b) The PM might be able to remain in the fairly large L4 "tadpole" region for millions of years, then orbit as a horseshoe object for tens of thousands, then be trapped in the L5 tadpole region, and so on. It is beyond the scope of this paper to work out such scenarios, especially since the possible migration of the major planets, chaos effects, and perturbations by other transient objects (including comets) can change the picture. Chaotic effects can even cause planetary collisions (Laskar & Gastineau 2009) on a long timescale! (c) If the eccentricity and inclination are right, Mikkola et al (2006) might handle the problem.

## Outline of the new theory of the Moon

The stages in the creation of our Moon as presented here are briefly outlined as a guide to the rest of this paper.

I. Formation of the PM at L4. This follows the BG approach except that the mass of the PM is about $M_\oplus/48$ - $M_\oplus/39$ (the latter is preferred), not $M_\oplus/10$. Breakout ensues *after* PM core formation. There is a range of allowable values, due to uncertainty about the amount mass lost to the Earth vs what ends up in the Moon itself, but we prefer 1:39.

II. The PM moves in horseshoe orbits approaching Earth. This again follows BG except that we calculate radial and tangential tidal forces and select an orbit allowing capture at a radius $a_0$ slightly outside the Roche limit. Interestingly, Horedt (1976) found orbits similar to those of BG but found capture, not collision, as I do.

We have calculated captures from hyperbolic orbits with velocity at infinity $V_\infty \sim 35$ to 180 m/s and impact parameter $\sim 60$ to 500 $R_\oplus$. For capture, the larger velocities must be matched with the smaller impact parameters. To a first approximation, we require the post-capture proto-lunar semi-major axis $a$ to be within a Hill sphere radius

$$a_{Hill} = AU \sqrt[3]{M_\oplus / (3M_\odot)} \qquad (1.1)$$

or about 234 Earth radii. On a more sophisticated level, the new position should lie within a surface such as shown in Fig. 5.4 (a), (b), or (c) or (d) of Roy (2005) defined by the Jacobi constant

$$C = x^2 + y^2 + \frac{2(1-\mu)}{r_1} + \frac{2\mu}{r_2}, \qquad (1.2)$$

where $\mu$ is the Earth/(Earth+Sun) mass ratio $\sim 3 \times 10^{-6}$, $x$ and $y$ the lunar coordinates and the radii $r_i$ are the distances from it to the Earth and Sun respectively, all in units of the Earth-Sun distance. When dissipation is included, C will not be constant, presumably increasing from a value less than 3 to one between $C_1$ and $C_2$, as described in Roy. In our case the dissipation is tidal, while for BG it is, in a sense, frictional, being due to random gravitational encounters. In the BG case one expects that there are also cases such that C *decreases*, because their analysis is more general than that of dynamical friction (Chandrasekhar 1949). We found the Jacobi constant to behave poorly (variations by several parts in a $10^6$) using the default machine precision in *Mathematica* ~15.95 and had to increase precision to 48 (decimal) digits to get conservation to parts in $10^9$.

Our orbits closely resemble those of Malcuit et al (1989), so only a few will be shown here. These authors favored an "intact capture" theory, so their incoming body had the mass of today's Moon. They noted that true capture required trapping within about 230 Earth radii, though they did not describe that as the Hill radius. They ignored the problem of the large iron core to be expected with intact capture. Malcuit (2011) goes further, asserting that the Moon formed near the orbit of Mercury, but nevertheless failed to have a very large iron core. Malcuit's (2011) paper is followed by one by Gott (2011), who again summarizes the GI theory.

Fig. (1) shows an orbit starting at L4, circling the Sun-Earth CM, and being captured within the Earth Hill sphere. The arrival at Earth can't be shown to scale in this figure but is at the left and is detailed a little later. We use the BG reference frame, with coordinates displaced to put the instantaneous Earth center at (0,0).

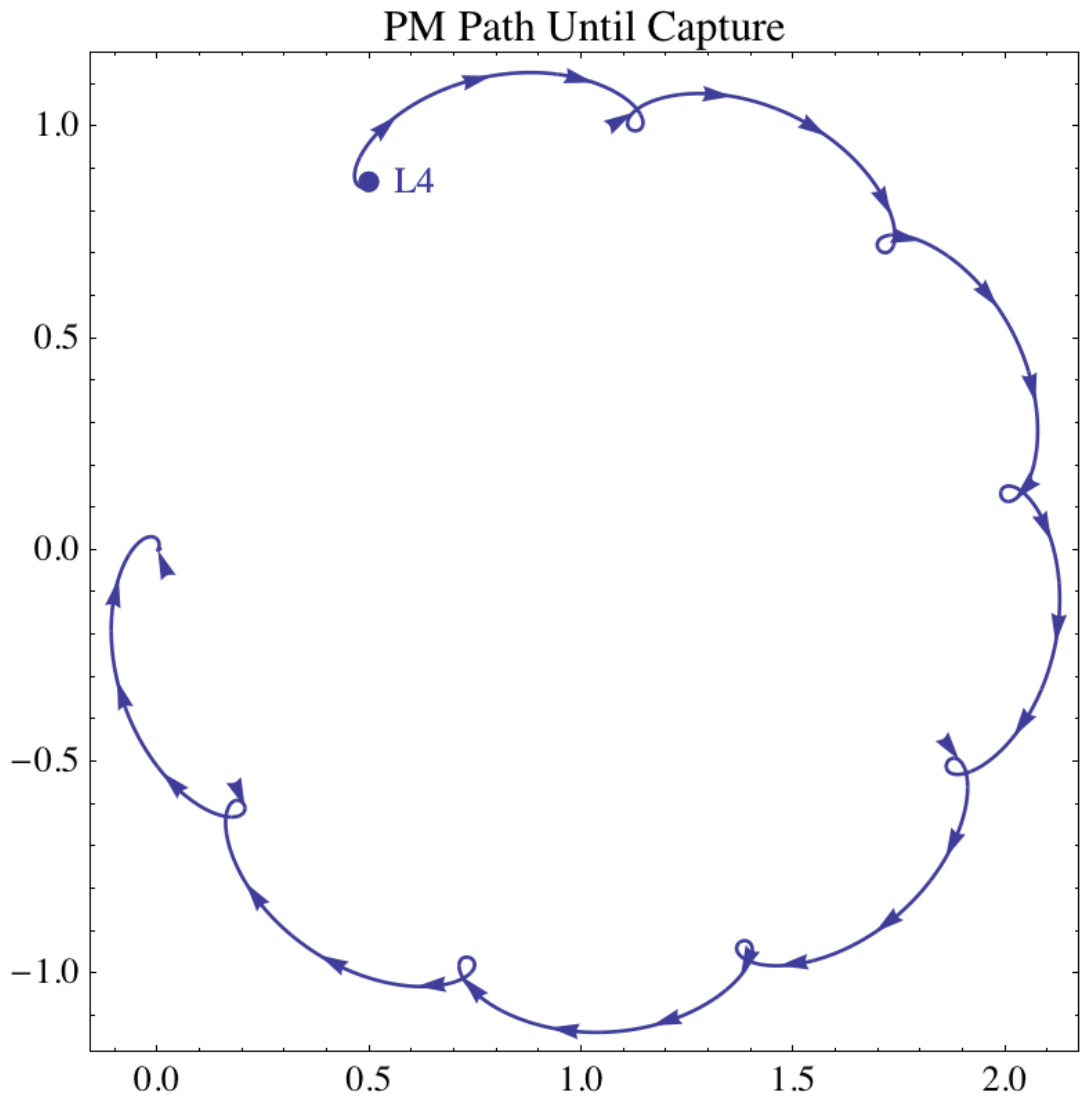

**Figure 1**. Path of the Proto-Moon (PM) after escape from L4 until Earth encounter. The Earth is at (0,0) and the Sun is at the center of the figure, almost (1,0). On this scale, capture details are too small to be shown, so the trajectory ends at Earth. Capture is due to tidal forces.

For comparison, to validate that we can produce results like those of BG when tidal forces are omitted, we exhibit Fig (2). The time range is from ~0 to 213.3 BG units (34 yr). Note that our value for the initial velocity at L4 is 0.0375931 BG units, which exceeds, as it should, their value 0.011 for which they state that horseshoe orbits morph into "creeping breakout" orbits (i.e. the orbits circle the Sun, not reversing just before Earth-encounter). With tidal forces included, experiments to attempt capture on passes by Earth later than the first (but not on the first) met with failure after many tries. It seems likely that capture happens on the first encounter or not at all, perhaps because the tidal forces perturb the orbit into larger miss distances on later passes, but perhaps also because we do not supply the tiny random perturbations of BG.

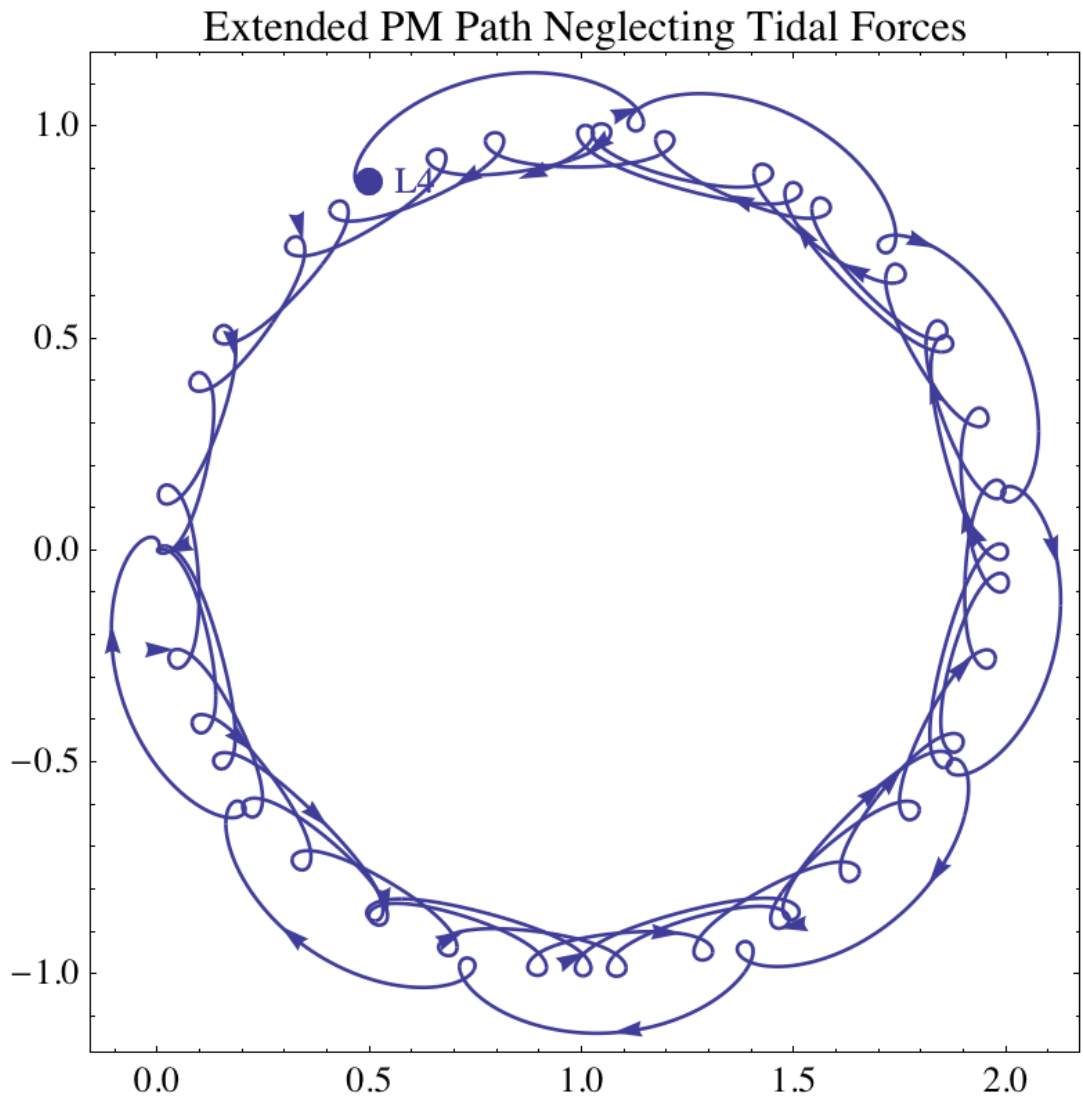

**Figure 2**. Path of the Proto-Moon (PM) after escape from L4 on the assumption of neglect of tidal forces. The starting conditions are the same as in Fig. (1). The trajectory can be continued quite far, but the loops overlap after a while.

IV. Approximate circularization of the proto-lunar orbit by tidal forces at $a_0 \sim 2.825$ Earth radii, just inside the co-rotation radius follows the capture. This sets the scene for Roche stripping, which actually begins at 2.713 Earth radii. The circularization is due to azimuthal tidal force being small at apogee, and very large at perigee, while there are also radial tidal forces tending to circularize the orbit. Fig. (3) shows the circularization of the same orbit as in Fig. (1), but in an Earth-centered frame. The coordinate system (after BG) is still one that rotates once a year, and the BG equations of motion, which account for centrifugal and Coriolis forces are used again, supplemented by tidal forces. The three concentric circles are the Earth surface, the Roche limit and the co-rotation radius in order of increasing size. Note that the closest approaches are early on; circularization increases the perigee distance as it decreases the apogee.

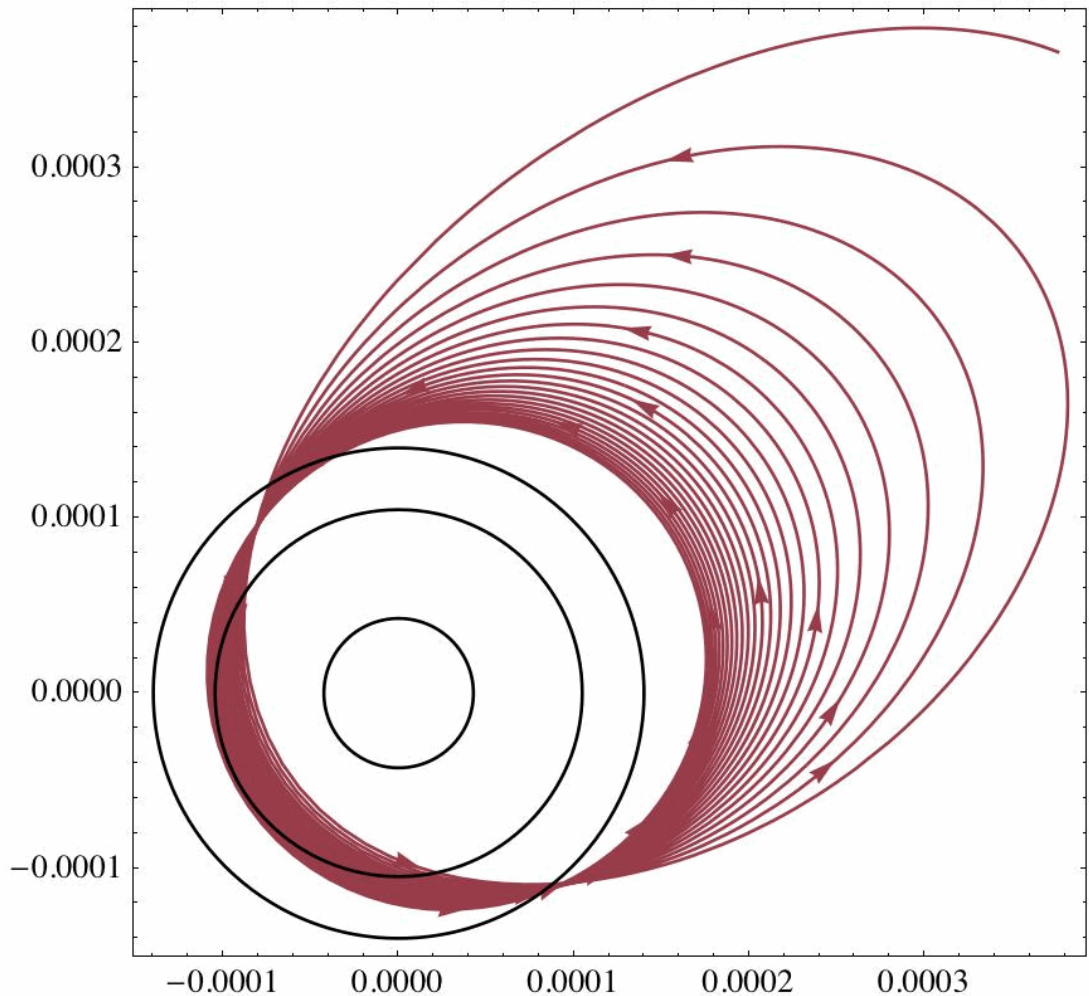

**Figure 3**. Path of the Proto-Moon (PM) during the early part of orbit circularization. The three concentric circles, in order of decreasing size, are the co-rotation or synchronous radius, the Roche limit, and the surface of the Earth.

The mid-late stages of convergence toward Earth are shown in Fig. (4). As in BG, r2 is the Earth-Moon distance in A.U. On the timescale shown, it might seem that the PM is settling just outside the co-rotation circle, but if the plot is continued much farther to the right, it dips down within that radius *and* the Roche limit! But with any reasonable horizontal scale, the radius variations become a blur. The horizontal scale of the oscillations in *r2* then degenerate until they are a smear.

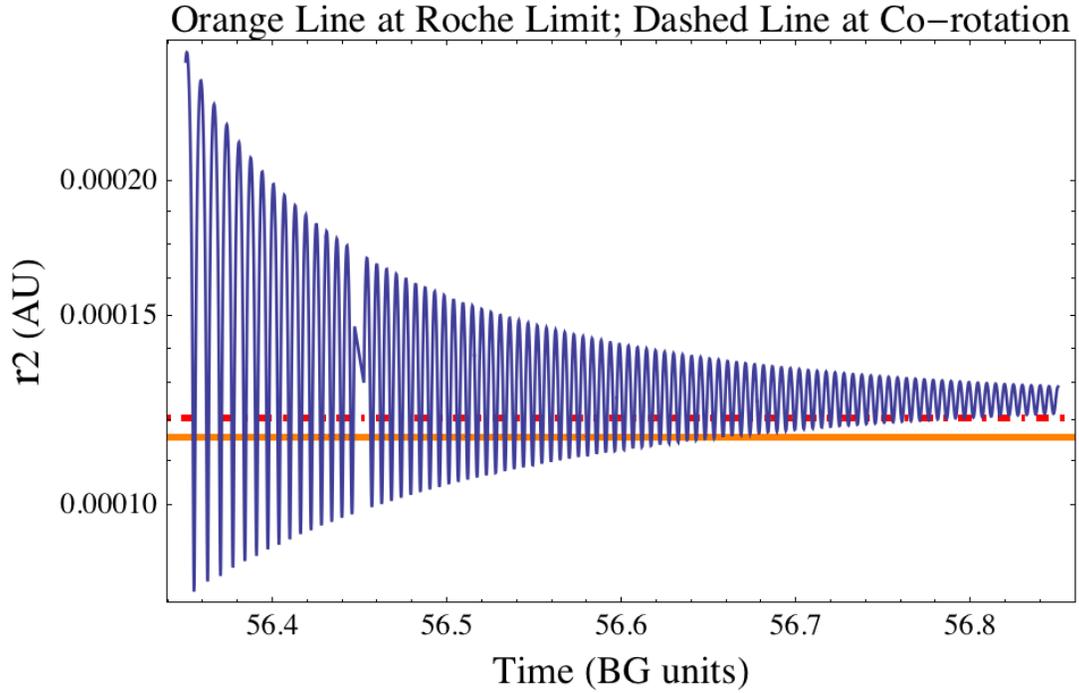

**Figure 4**. The distance from Earth center to the center of the (PM) during the early to mid part of orbit circularization. If continued to the right (increasing time), the path will slump down within the Roche limit. Time in BG units of $2\pi\ \text{yr}^{-1}$.

Since osculating elements are a poor representation for a rapidly changing orbit, we exhibit in Fig. (5) instead of the osculating eccentricity, a graph of the pseudo-eccentricity, which we define as

$$\varepsilon_{pseudo} = \frac{r_{apogee} - r_{perigee}}{r_{apogee} + r_{perigee}} \tag{1.3}$$

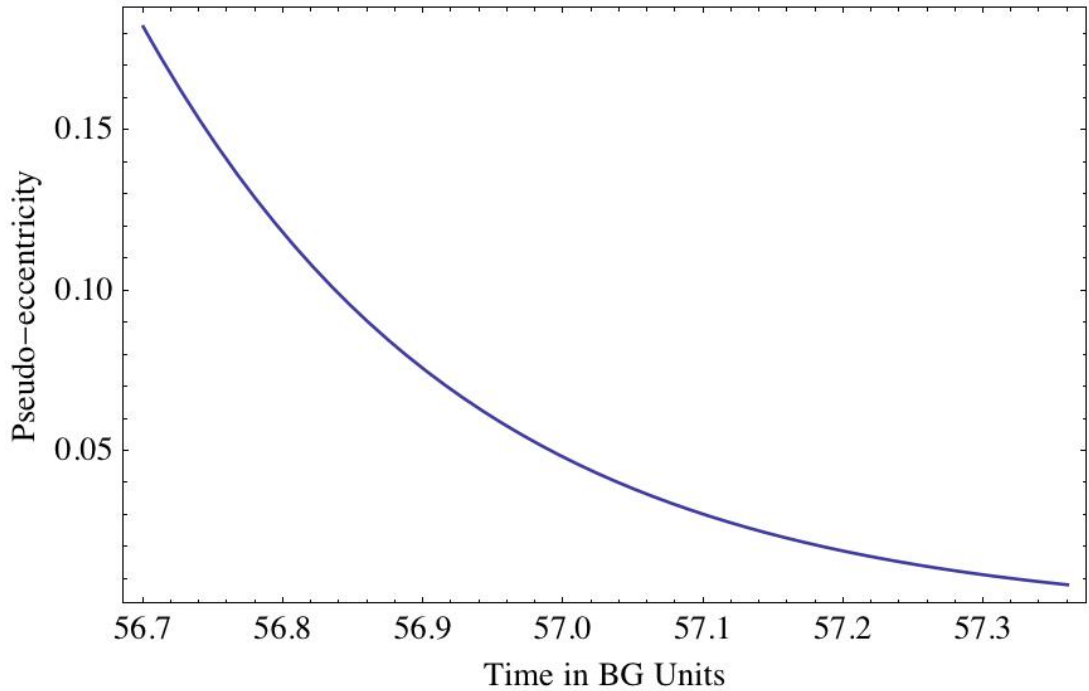

**Figure 5**. Decrease of the pseudo-eccentricity (see text). Time is in BG units.

III. The PM orbit approaches Earth due to tangential tidal force, because the orbital period of the PM is a little less than a sol (primordial day); i.e. it is inside the co-rotation radius. This process is shown in Fig. (6). The PM passes the "point of no return" after one entry and exit from the Roche limit. (Recall that Holsapple and Michel (2006, 2008) showed that entry within the Roche limit for a short time does not tear the object apart). The orbit was followed numerically in more detail than we can show here and energy loss in the Moon (heating) was evaluated. The following Figure was generated by isolating the close approaches using a plot like Fig. (10) of BG, but with more cycles, as in their Fig. (16). Their Fig. (10) does not show where Theia entered the co-rotation radius at dimensionless distance ~ 0.000119, nor the Roche limit at dimensionless distance ~ 0.0001, as they are concerned with collision, which we avoid, for our less massive "arrival," not with capture. We have verified in our "capture" cases that the closest approach always exceeded by a fair margin the distance $5.66 \times 10^{-5}$ AU, which is the "contact" or collision radius for the PM and the Earth in BG units. That radius lies hardly more than twice the width of the heavy dashed line above that line in Fig (6). The initial descent is just due to solar and Earth gravity, and after the Hill radius mostly to Earth gravity and inertia (our aim was good)

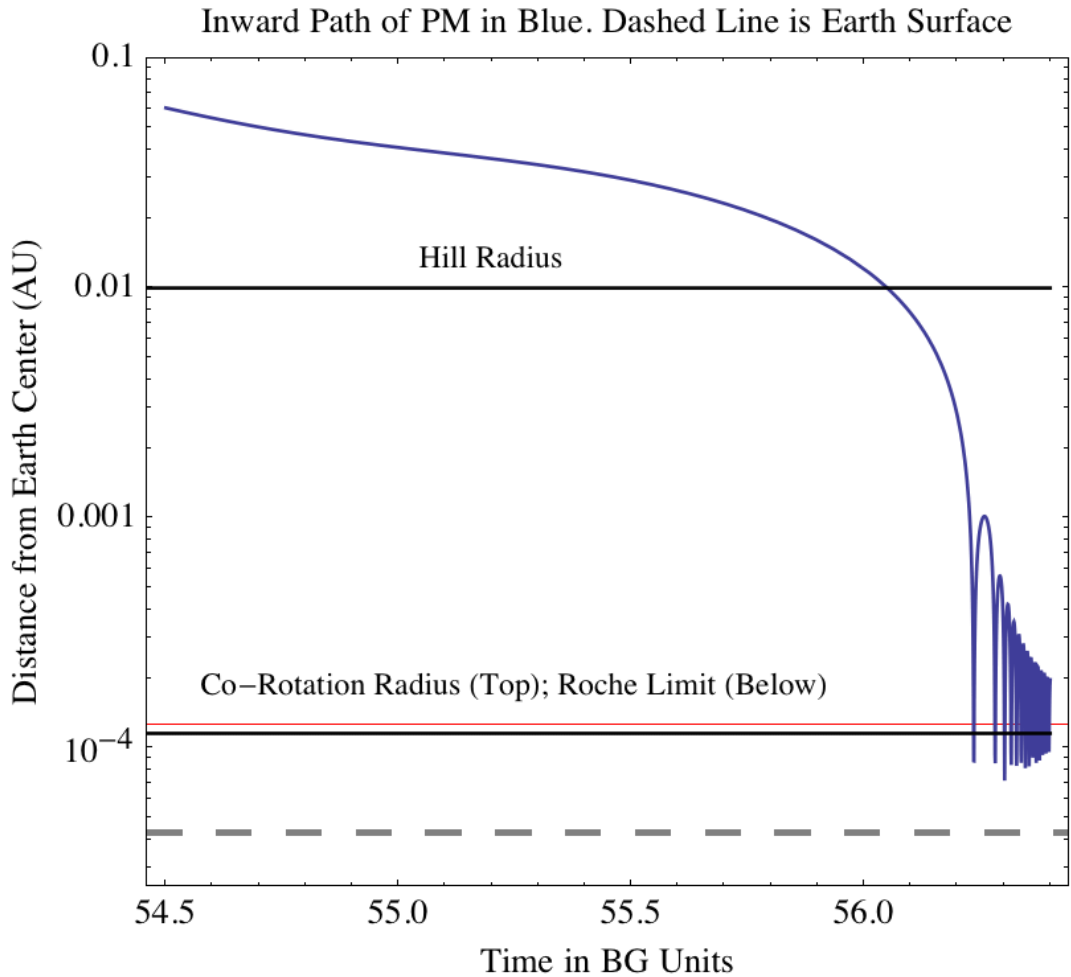

**Figure 6**. The distance from Earth center to the center of the (PM) during the capture process. The oscillations across the Roche Limit are real but damp later as the orbit circularizes.

V. A succession of Roche limits is traversed inwards, in each case the outermost layer of the PM being stripped into a ring (actually a disk would exist, but it is modeled as many rings – O72 imagined six.) The remaining object is the "partially stripped proto-Moon", which we refer to as the "Hull." The gradualness of the stripping is caused by the steady increase in mean density of the Hull, a maximum when all the rock is gone and the iron core remains. The Hull's surface gravity also tends to decrease with radius for the most part. As each rock layer is stripped, shepherding from the remaining Hull drives much of it radially outwards; much of the material is demonstrated to collect well outside the Roche zone at ~ 3 Earth radii. Some of it is driven in to strike the Earth. The time for the Hull to go from 2.7 $R_E$ to 2.24$R_E$ where all the rock is gone in only ~400 - 700 d, where "d" is the present day of 86400 s. The different timescales result from different assumptions on the Earth response times to tidal forces (see below).

Allowance is made for possible re-accretion and re-stripping. Re-accretion is actually a bonus in this theory because the Moon is obviously not "exactly" turned inside-out; re-accretion allows more mixing of stripped particles than would occur by mixing within the disk. Some re-ordering of layers will also occur as the Moon re-forms in the region 3.6 - 3.8 Earth radii. The evolution of the stripped material was followed in part with equations from Goldreich & Tremaine (1980), but, for simplicity, conservation of mass and angular momentum were used to find the final angular-momentum-weighted mean radius 3.8 $R_E$. Typical rock particle sizes, if estimated by equating the total particle surface energy to that dissipated in stripping could be of order 10 microns. If we use instead equations in Holsapple and Michel (2008) or Asphaug and Benz (1996) we find ~km or larger sizes. This problem merits further study. During the stripping period and the re-constitution period (below, Step VI), the comminuted rock particles are exposed to near vacuum and eventually to some heating by the former core (ball of Iron or Iron-Nickel or Iron-Sulfur). This allows escape of volatile elements, such as Ga, I, Na, K, Zn, Pb, and In as vapor, unimpeded by gas (Taylor 1986,Yin 2005). In the popular GI theory, much of the material that is to condense into the Moon is gas. Although turbulent diffusion might assist volatiles to escape, molecular diffusion is far, far too slow, so the loss of volatiles is rather a puzzle (Humayun & Cassen, in Canup & Righter 2005); Schmitt (2006).

VI. There is a sudden jump in density from ~ 3350 kg st$^{-1}$ for rock (see below) to ~7200 kg st$^{-1}$ when the Roche stripping reaches the PM core boundary, the last bits of rock having been stripped at $a$ ~ 2.233 $R_E$. The stripping process then acts on the PM core, which has reached the Roche limit for Iron. (The PM core is assumed to be iron-nickel alloy with perhaps sulfur, as contrasted with the present lunar core, which could be rich in silicates or Titanium). During the phase of core disintegration, the hot and possibly molten remaining core continues to drive the rings or disk of rock outwards, while heating the particles in the nearer rock crumbs. The iron disk or ring could persist for a long time, save that the rock layers have collected at ~ 3.3 to 3.8 $R_E$ where they merrily have re-formed into our Moon. This process is facilitated by the shepherding of the iron ball (PM core) as it descends below $a$ ~ 2.233 $R_E$. That shepherding is expected to ensure that the rock disk collects into a single Moon, avoiding the previously identified problem of multiple moon formation (Canup & Esposito 1996, Ida et al 1997, Kokubo et al 2000a,b). The radial ring velocities are large, of order 3 - 7 km s$^{-1}$. The radial ring speed tends to be about 1.6 times the orbital speed of ring particles, so they travel in loosely wound spirals, while the Hull travels a more tightly wound spiral downwards. For the interaction of the rings and the Hull we use

$$\frac{1}{a_{sat}}\left[\frac{da_{sat}}{dt}\right]_{ring} = 0.798 \frac{M_{sat} M_{ring}}{M_\oplus^2} \left(\frac{a_{sat}}{a_{sat}-r}\right)^4 \Omega \, sign(a_{sat}-r) \qquad (1.4)$$

and

$$\dot{r} = -\dot{a}_{sat} \frac{M_{sat}}{M_{ring}} \sqrt{\frac{r}{a_{sat}}} = 0.798 a_{sat} \sqrt{\frac{r}{a_{sat}}} \frac{M_{sat}^2}{M_\oplus^2} \left(\frac{a_{sat}}{a_{sat}-r}\right)^4 \Omega \qquad (1.5)$$

from Goldreich and Tremaine (1980), where the "satellite" is the Hull. Harris and Ward (1982) produce similar results with a coefficient somewhat less than 0.798 via heuristic arguments. In the present context, the subscript "sat" refers to the PM or, more accurately, its partially stripped Hull. There is no singularity in the denominator, because the ring distance *r* from Earth center cannot be closer to the orbital radius $a_{sat}$ of the Hull than the radius of the latter.

VII. Reverse shepherding now occurs, with the roles of the Hull or iron ball and rings interchanged. The newly formed Moon drives the iron ring and any portion of the stripped rock that had lain within the orbit of the PM as it lost mass down towards Earth, sapping its angular momentum and hastening the Moon's departure into tidal drift away from relatively low Earth orbit. The inward forcing of the left-over items is also described in Kokubo, Ida and Makino (2000)'s Step 7. The iron (PM's core) will be deposited on the surface of the Earth. Although its mass, ~ 4-5 × $10^{22}$ kg is considerable, it is many times *less* than the mass ~ 2 × $10^{23}$ kg of the core of the Mars-mass object Theia usually considered to have entered the Earth at a nearly *vertical* angle. Because it strikes obliquely, as a belt of iron and rock turned to vapor, its trace might be detected in Earth surface rocks such as may have survived from the lunar formation event. Indeed, it could explain the "late veneer" of chondritic material found by Wood et al (2006), and O'Neill (1991). The inner iron ring and rock disk arrive at the Earth's equator (of date) at LEO (low Earth orbit) speed ~ 7.9 km s$^{-1}$ and virtually zero angle of incidence. Our calculations produce a veneer of $7.77 \times 10^{22}$ kg rock and iron total, of which $4.8 \times 10^{22}$ kg is iron comprising the PM core. This is somewhat larger than the figure $0.7 - 2.7 \times 10^{22}$ kg determined from a few trace elements by Dauphas & Marty (2002), but it is hoped that our numbers are acceptable. Boyet and Carlson (2007) suggest that the superchondritic Sm/Nd ratio they find in some lunar samples implies that these rocks came from Earth, possibly near Isua, Greenland. Our position would be that it is more likely that the Greenland rocks came from the Moon! Our number for mass accretion is about four times the figure given by Bottke et al (2010). Their estimate was conditioned both by HSE content and estimates of impactor numbers; the latter would tend to lower the over-all mass estimate as it is difficult to provide a source for so many impactors, as well as to justify the lower rates of late accretion on Mars and the Moon. In fact, Bottke et al find it "curious on several levels" that the Moon seemingly suffered less late accretion than Earth. They also need (their footnote 31) to prevent the accreted material from entering Earth's core. In our case, the reduced impacts on the Moon are natural, because late accretion to Earth was material *from* the proto-Moon. Also, the lack of entry to Earth's core was due to the very flat angle of incidence. Unfortunately, the heat generated on impact would be $2.4 \times 10^{27}$ kJ, enough to vaporize the $7.65 \times 10^{23}$ kg of rock in a shallow, 300 km deep "magma ocean" (Solomotov 2000). Even these numbers ignore the sensible heat in the veneer, which is assumed fairly cool due to radiation to space during its descent. The arrival of the veneer would have sent the primordial atmosphere off into space. The amount of energy and heating described here is, of course, much less than that in the GI theory, wherein most of the Earth ostensibly melted. Although our theory causes a more modest disturbance of the early Earth than the GI, we would still like to look for mitigating factors. One is that the rock would arrive gradually, although the time period would be only ~ 1 hour. The iron would arrive a little later. This allows for a little cooling. It would be possible to reduce the heating (so as to melt the rock in the "magma ocean" and not vaporize it) by assuming a somewhat smaller PM mass, though we can't go too near $M_E/54$, or else by allowing the escape of some mass from the system, which we have set up to conserve mass in order to get a well-confined theory. The GI theory involves the loss of perhaps two thirds the mass of Theia. At present we will not further pursue avenues to lower the energy deposition. Note that Touboul et al (2012) conclude their analysis with the remark that it is unlikely that the whole Earth melted by the event that formed the Moon if that occurred more than 2.8 Ga ago, as it does in the GI theory. Also, Walter et al (2000) state that "there is no independent evidence that a magma ocean ever existed on Earth," which supports our position.

## Was the Moon Turned Inside Out?

According to scenarios for lunar formation (O72), and by the author (Noerdinger 2010), the Moon was formed by multi-stage disintegrative capture, a process in which capture occurred into a descending (month shorter than day) orbit just outside the Roche limit (RL), followed by the proto-Moon (PM)'s being stripped, layer by layer, into a disk. The outermost disk portions would have been low density REE (Rare Earth Element)-rich rock from the crust and upper mantle of the PM, while the rings stripped from the Hull at smaller geocentric radius $a$ would have come from the lower mantle. The reason for this is that the outer rings (idealized portions of a disk) were not only stripped first (at large semi-major axis $a$) but were driven outwards by torques from an almost full-mass PM. Rings stripped later started their outward trip nearer to Earth and were propelled by a less massive, partly-stripped PM Hull.

So long as a ring of rock matter (really a portion of the disk) is inside the RL, it cannot condense into a single self-gravitating body, but the portions outside the RL, which have been rapidly driven there by the tidal impulse of the remaining portion of the PM inside those portions, can, indeed re-form into what will become our Moon as in Kokubo et al (2000) and Ida et al (1997). The inner-mantle portions of the PM arrive outside the RL later, and will accrete to form an almost inside-out Moon, while the iron PM core, being fully stripped at $2.233\ R_E$, forms an iron disk which descends to Earth, driven down by the tidal action of the newly formed Moon.

Of course, we cannot have a clean inversion of the original layers, because of turbulence and random processes. Yet the scenario stands in broad outline. The density inversion in the newly re-accreted Moon resulting from the inside-out reprocessing is unstable. Lavas can also rise by evolving gas in the usual way. There may be a magma ocean, forming a plagioclase crust, but in that case the survival of the magnetized rocks is difficult to explain. Instead, the plagioclase may have ascended in numerous diapirs (Longhi and Ashwal 1985).

The small iron (or other) core of the Moon today would be a secondary one, confirming the abstract of Neal et al (1999). Geological studies of returned Moon rocks, e.g. Norman et al (2010) often find discordant ages among adjacent rocks or clasts. The disagreements can be attributed to the resetting of ages on impact, or the strewing of impact ejectae from regions near and far, often Imbrium. We would contend that the older samples are survivors from the formation and differentiation of the PM, while ages < 3.8 Ga date from the GE.

## Heating of the Proto-Moon During Capture and Stripping

Gerstenkorn (1955, 1969), Singer (1968) and Öpik (1972) considered the heating of a moon during Earth capture. Such heating is a substantial problem for the theory presented here, because many lunar rocks and soils are ~4.4 to 4.6 Ga old yet the capture of the PM is herein claimed to have occurred ~3.8 - 4.13 Ga ago. The older age limit is mainly due to the age of Troctolite 76535 (just one rock!), which has a magnetization that I presume to have survived the GE event. Assuming that this claim is correct, the rock should also not have been heated to more than 1123K, its Curie point, after it was magnetized (Garrick-Bethell et al 2009). Referring to Eqs. (2.14) - (2.15), we see that the only term causing substantial lunar heating is proportional to $k_{2m}$. That heating occurs when the radial motion is large, which is when the geocentric lunar semi-major axis $a$ is near the co-rotation radius. Although Gerstenkorn (1969) emphasizes the radial tidal forces in facilitating capture, and thus finds severe heating, with whole-Moon melting, I claim that capture and orbit circularization are as much or more due to ordinary tidal force $F_\perp$ *normal* to the Earth-PM radius vector *r* as to radial forces; the normal or tangential tidal forces speed up the Earth and slow the Moon near perigee, while being weak but of opposite effect near apogee. This situation is emphasized in Singer's Fig. (1). The force $F_\perp$ is inversely proportional to $Q$, the Earth quality factor, which we take to be ~9 for the primitive Earth. Values ~1 or less invalidate the usual tidal equations (Murray and Dermott 1999, Meyer et al 2011). The Love number $k_{2m}$ for the PM plays an important role in the heating by radial tides. We assume it is $k_{2m} = 0.055$, which is a bit larger than that of Peale and Cassen (1978). The larger value is due to our assumption of more sudden and severe heating, and is, in a way, self-defeating; since we do not want too much heat we might plump for a value ~0.02, but parameters were adjusted to effect capture more than to control heating. A fairly low value of $Q$ is needed to allow the azimuthal tidal forces due to Earth to dominate the other tidal forces. It is a difficult task, sensitive to $Q$, the Earth relaxation time $\tau_E \sim 1000s$, $k_{2m}$, and the initial values of the position and velocity very near those of L4 itself, to estimate the heating of the PM There is a sudden cutoff of whole-PM heating as Roche stripping begins, by which time the orbit is essentially circular. Our calculations, by integrating the part of the radial force on the PM due to its own deformation times its radial (toward/away from Earth) velocity can lead to temperature rises as small as 250 K or as much as 10,000 K, assuming a specific heat of 0.81 kJ/K, as for rock. We tune parameters to keep the temperature before stripping at the Roche radius below 1,200 K. Up to that time cooling by radiation is small, but as fragments break off they can cool rapidly.

## Heating of the Moon During and After re-Formation

In a manner similar to that described in Kokubo et al (2000) and Ida et al (1997), the Moon accretes from a disk, but in our case the disk is made up of cool or cold rock particles, *outside* the Roche limit by a fair margin. The accretional heating of the Moon if it forms at ~3.8 Earth radii by simple accretion as a single object would melt it, and so would generate the magma ocean, as is generally accepted. The existence of a lunar magma ocean has, however, been challenged by Gross et al's (2012) analysis of lunar meteorites. These authors state that the meteorites, which represent a wide area on the Moon, demonstrate that ferroan anorthosites are peculiar to the Imbrian basin! Equating the accretional energy based on the escape velocity of the partly accreted Moon and loss by black-body radiation with emissivity 0.85 - 0.95, layer by layer, would generate enough heat to melt basalt at ~1050K or more and even Dunite at 2163K unless one stretched the accretion period to a ridiculously long 5000-6000 yr period to prevent melting the Basalt, or at least 300 yr to save the Dunite. On the other hand, *hierarchical* accretion (Kokubo et al 1998, 2000; Asphaug & Benz 1996) could lead to enhanced cooling during the re-accretion process, due to the larger surface-to-mass ratio in multiple components, so that the melting could be marginal or not happen at all. This would, desirably, preserve rocks magnetized by a dynamo in the PM (see below). Unfortunately, Kokubo et al and Asphaug & Benz did not calculate temperatures, and the work, in any case, was inside or very near the Roche limit, not at 3.8 Earth radii, so it would be difficult to adapt to our case. This will be left mostly an open issue, although we have put some thought into keeping the accreting layers cool this way. If the single accreting object is replaced by N identical ones with the same total mass, the surface goes up like $N^{1/3}$. Assuming 10,000 objects accreting and cooling as they accrete, and ignoring the final energy of merging when they form into one Moon, we get only a factor ~1/25 shortening of the allowable accretion time if we wish to avoid melting, say, basalt. That would be ~200 yr. In that time, the pieces would move outward a negligible distance from Earth center, because the small masses mean small tidal forces. The more massive ones would move the most. This solution seems a plausible way to save the magnetic rocks and deserves more study. Furthermore, since the accreting pieces are co-orbiting near the partly accreted Moon, they might strike at less than its escape velocity.

We have admitted that accretion as a single object would cause a whole-body temperature rise of ~2000 K, more than enough to melt basalt, but some heat would be lost by radiation as the heating is mostly at the surface during the whole re-accretion process. Pritchard and Stevenson (2000) obtain much higher temperatures from accretion than we, due to their assumption of a hot disk from a giant impact, and substantial eccentricity and inclination of the disk particles' orbits. Our disk would be comprised of cold rock (perhaps 100 K rather than their 500 K) and the inclinations would be small. Eccentricities might also be damped by interparticle collision. Pritchard, and Stevenson's lunar mantle temperatures after accretion become very high as they assume a hot initial state and primordial U and radioactive K abundances, while in our Case II another 260 Ma had passed so that the radioactive K is a bit depleted. In our scenario, the initial lunar material has cooled in space. In any case, the unconsolidated mantle rock layer is a challenge for theories that produce melting of the whole Moon. In Case III, the present Moon is formed ~760 Ma after the Earth. By then, $^{40}$K is almost half gone and $^{235}$U more than half, so radioactive heating of the interior is more modest than in the cases studied so far (Elkins-Tanton et al, 2011, Pahlevan & Stevenson 2007). A hot upper mantle, as we find from accretional heating, and cooler deep interior are in accord with the finding by Watters et al (2012) of recently formed graben indicating recent expansion of the

Moon, a finding which also is discordant with the GI theory, as Watters et al disfavor a totally molten original Moon.

Before dwelling further on our disintegrative capture theory, we present a table similar to that in Wood (1986). Table 1 shows a comparison of various theories of the origin of the Moon.

Table 1.
Intertheory Comparison

## Requirements:

| | | Numerical/Data | | Comparison |
|---|---|---|---|---|
| A. | Total Angular momentum | 3.50E+34 | kg m^2/s | Earth 1.16E+34 |
| B. | Produce one and only one Moon | 1 | | |
| C. | Lunar Mass | 7.35E+22 | kg | |
| D. | Moon to have tiny iron core | 1 to 3% of the Moon's mass | | Earth: 32% |
| E. | Oxygen 18O/16O/17O | same in Earth and Moon | | meteorites |
| F. | Depletion of Volatiles | selective | | |
| G. | Magma Ocean | probably existed on Moon | | Troctolite 76535 |
| H. | Magnetic Field | dynamo req'd | | basalt 10020 |
| J. | Feasible | Yes/No | | |

Scorecard (mostly adapted from John Wood in "Origin of the Moon" - LPI - available at NASA
(W) with an entry indicates a Wood judgement not echoed here

<span style="background:lightblue">　</span> means my judgement　<span style="background:red">　</span> means "fatal"　<span style="background:lightgreen">　</span>　　　　　means not in Wood

| Scenario-> | Intact Capture | Co-accretion | Earth Fission | Big Whack | Disintegrative Capture | Capture+Roche |
|---|---|---|---|---|---|---|
| A. AngMomentum | C (W) | F | F | B | C (W) | B |
| B. (1 Moon) | A | B | A | D | C | A |
| C. Mass | B | B | D | I (W) | B | B |
| D. Small Core | F | D | A | B | B | A |
| E. Oxygen Iso's | B | A | A | B | B | B |
| F. - Volatiles lost | C | C | B | B (W) | C (W) | B |
| G. Magma Ocean | D | C | A | A | B | B |
| H. Magnetic Field | D | D | D | C | D | A |
| J. Plausible | D- (W) | | F (W) | B | F (W) | TBD |
| Original Proponents | Gerstenkorn | Harris, Kaula | G. H. Darwin | Hartmann&Davis | Ruskol (1960-65) | P. Noerdlinger |
| Later Proponents | Conway | | | Cameron&Ward Canup Belbruno&Gott | Mitler (1975) | |

In addition, we present a brief Table comparing our Cases I, II, and III

Table 2
What our Cases May Explain

| Item                Case | I                 | II                | III        |
|--------------------------|-------------------|-------------------|------------|
| Troctolite 76535 - **B** | unexplained       | explained         | explained  |
| Lava 10020 - **B**       | unexplained       | unexplained       | explained  |
| Lunar distance now       | why not larger    | why not larger    | maybe OK   |
| Oldest structures        | OK                | questionable      | problematic|
| find veneer belt today   | almost impossible | almost impossible | difficult  |

The boldface **B** refers to magnetization. The concept of finding or locating the belt of iron and rock comprising the Late Veneer, which was deposited in a narrow strip along the Earth's equator when the Moon formed, requires tracing continental drift back to that epoch. If the Moon is older than the beginning of the Archean age (which we take to start right after the Hadean), the task looks hopeless. It is hoped that GRAIL may resolve the ages of the oldest lunar structures, as the Moon has to be older than the oldest of those.

**Origin of the Maria**

The maria are recognized to have been formed by collisions early in the lunar history, each collision followed by upwelling of lava. The author has had the impression that the gravity anomalies (mascons) have been attributed mainly to that lava, the projectile (we use the term instead of "impactor" for obvious reasons) having possessed smaller mass, and being perhaps more deeply buried. There has been no explanation for the apparent time coincidence between the formation of the Maria and the late veneer (on Earth), other than that both could be attributed to a rain of projectiles suddenly issuing from the asteroid belt. There has also been reported a "late heavy bombardment," smaller than the veneer (Chapman et al 2007), but Hartmann et al analysis showing that the apparent time-peaking of the LHB is an artifact.

According to Ryder et al (2000) the veneer could have occurred as early as 4.40 Ga or as late as 3.8 Ga. If the rocks dated at 4.360 Ga by Borg et al (2011) are markers for the lunar creation sequence described here, then that is our age for the Moon (the actual capture and stripping process takes days to weeks). The concept that the projectiles that caused the maria came from the outer solar system fails to explain why the maria are on the near side of the Moon. It does provide high impact velocities, which allow their formation by relatively low-mass projectiles. Baldwin (1971) has asserted that such high velocities lead to much of the projectile being cast back into space, again a reason to expect that the mascons are due to lava. One possibility is that the impacts were due to the arrival of the last pieces of rock formerly attached to the lunar core, now driven out by precisely the action of that core just before it descended to its Roche limit and turned into a ring or radially narrow disk. It is assumed that the Moon has coalesced at ~3.8 $R_E$ by the time that the last chunks of rock arrive; perhaps they were considerably denser than the previously stripped material, and so hung on to the core a bit longer. This explains the near-side concentration of the maria, but, assuming rock projectiles, one finds that the masses are comparable to or larger than those of the lava flows. I have estimated the projectile masses using Eq.(7c) of Grieve & Cintala (1998) (GC). I assumed arrival at escape velocity 2.4 km s$^{-1}$, crust density 2900 kg m$^{-3}$, and projectile density 3400 kg m$^{-3}$. I took the value of the transition diameter as 18.7 km as GC provide, meaning that all ours are all larger. Table1 below shows the assumed crater radii and the projectile masses from the aforementioned equation. I also experimented with iron density but got much larger masses and, anyway, it seems unlikely that iron would split off the core.

| Name | Diameter (km) | Projectile Mass (lunar mass unit) |
|---|---|---|
| Imbrium | 1160 | 0.001 |
| Humorum | 820 | $3.2 \times 10^{-4}$ |
| Orientale | 930 | 0.001 |
| Grimaldi | 430 | $3.9 \times 10^{-5}$ |
| Schrödinger | 320 | $1.5 \times 10^{-5}$ |

The mass values are probably good to only one significant figure, not only because the input numbers are estimates, but because if Eq. (7c) of GC is compared to their Eq.(7a), one sees that the exponents were converted to not better than two significant figures. In summary, I find that the theory presented here provides a natural way to create the maria, all on the side near Earth, and at virtually the same moment that the late veneer hits Earth. It is possible that the GRAIL mission (Zuber et al 2011) could detect the remains of such large projectiles, though the rock density is probably close to that of the lower crust. The impactor masses are rather large for Hull fragments, but may be plausible.

**Initialization**

We match the total angular momentum $h_{Total}$ of the Earth-PM system to its present value plus a 1 per cent increment to compensate for loss to solar tides in the last 4.5 Ga (Canup, Ward & Cameron 2001), and an allowance for mass and angular momentum lost to the system in late stages. Thus $h_{Total} = 3.5 \times 10^{34}$ kg m$^2$ s$^{-1}$. Note that it is exactly the extra mass in the iron core of the PM that sets up our later inward migration process, which will strip the rock, because the Earth has to revolve slower than in an ordinary "capture" theory. It is often said that the Earth-Moon system has such a large angular momentum that it is difficult to explain. The situation was put in context by Fahlman and Anand (1969), who showed that in the big picture, the system is nearly on a universal curve that is yet to be explained. Their Fig. (1) follows as our Fig. (7). On it the Earth-Moon system seems to fall right on the trend line of angular momentum vs mass, but it's really a factor 10 high - a negligible difference within the "big picture."

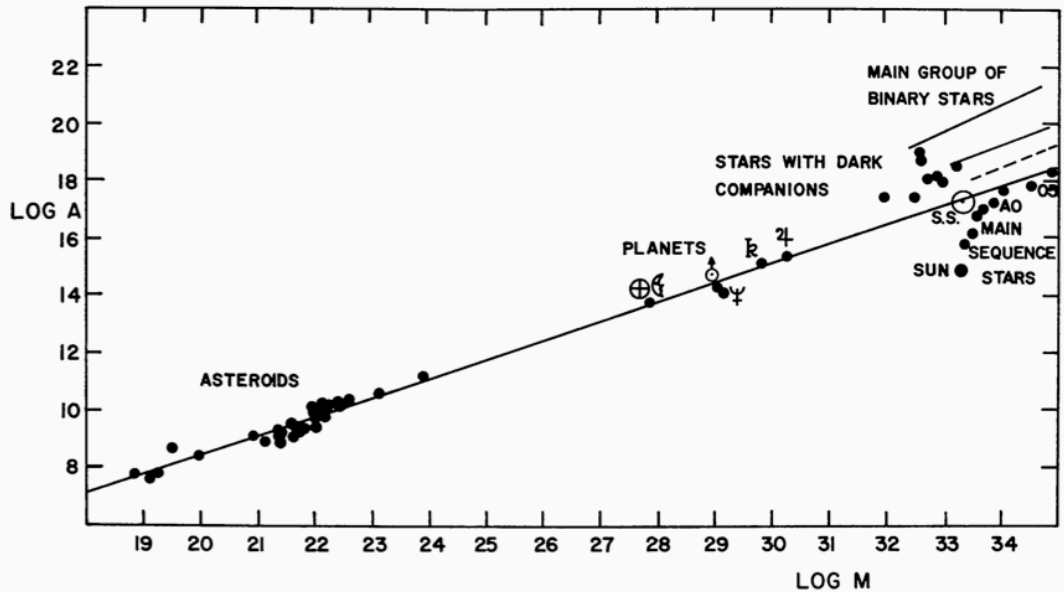

FIG. 1—The distribution of angular momentum per unit mass, $A$, with mass, $M$. The dashed line indicates the lower limit for the visual binaries. From Fahlman and Anand

**Figure 7**. The grand picture according to Fahlman and Anand - original caption is included.

## Proto-Lunar Interior Model

The PM will have an iron-nickel core (with traces of other elements), which we model here for simplicity as pure iron. The mantle will be rock. Crust is considered part of the mantle. The model was modified to include compressional effects to some extent in order that the Roche stripping process could proceed in an orderly way; however the dependence of surface gravity on radius, even for a two-component core/mantle model with constant density in each part is sufficient for most purposes to relate radius with orbital radius as the PM is stripped. Therefore we did not use the model with density variation within the rock. At the very end, however, it would be worthwhile to model the disintegration (stripping) of the iron core more accurately, because we need it to drive the outer rings of rock well beyond the Roche limit for rock before the core disintegrated into a ring, which is forced down into the Earth by the action of the newly re-formed Moon. This modeling with density varying in the core will be left as a future project.

We assume a base level density of iron at pressure $p \sim 0$ by $\rho_0 \sim 7200$ (a little high to allow for Nickel; and ignoring possible Sulfur). This density "guesstimate" is much less than that of Earth's core, because the pressure is lower. It is a bit larger than the estimate of Rivoldini et al (2011) for Mars' core density, 6600 kg st$^{-1}$. We do not consider the difference to be highly significant; Mars is thought to be a planet with "arrested development," and having a core only ~24% by mass, while our PM is assumed to have 32%. Our assumed PM rock mantle has a density 3272 kg st$^{-1}$.

The approximate structure of the core was calculated by iteration. Assuming a density $\rho_0$ = 7200, and ignoring the already-stripped mantle, the pressure at radius $r$ is

$$p(r) = \frac{2\pi G \rho_0^2}{3}\left(r_{surface}^2 - r^2\right) \quad (1.6)$$

The surface radius $r_{surface}$ will, of course, decrease at the iron ball is Roche-stripped. This is accounted for in the analysis programs. The bulk modulus of iron is about $K_b = 170$ GPa (Environmental Chemistry 2008). The actual density will then be

$$\rho(r) = \rho_0\left[1 + p(r)/K_b\right] \quad (1.7)$$

The core structure would cause the ring into which it disintegrates to have a finite width, but we have not modeled that feature. Our model for the structure as reflected in the surface gravity as a function of radius, which is also the internal gravity acceleration, since a spherical shell produces no internal field, is shown in Fig. (8).

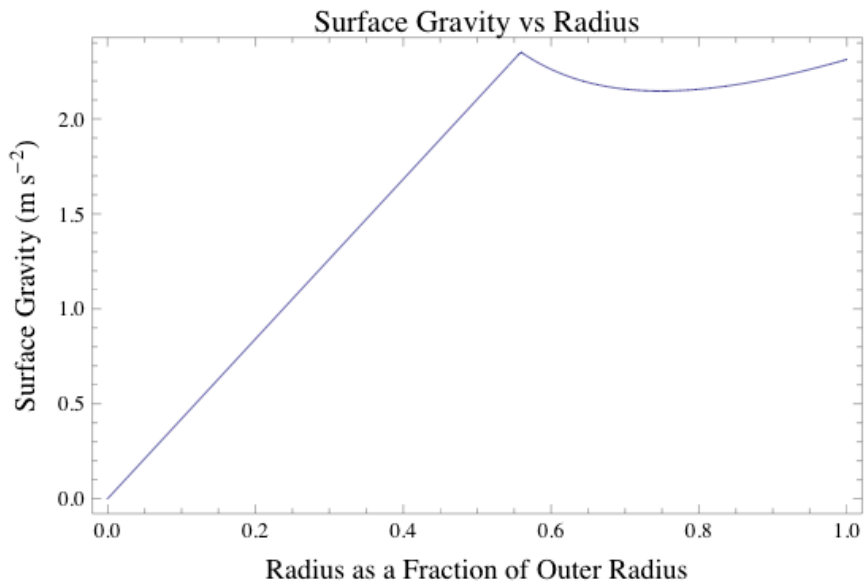

**Figure 8**. The gravitational acceleration in the PM versus radius, before or during stripping. The straight-line portion is within the core.

## Tidal forces and capture processes

BG did not consider the possibility of tidal capture of their "PM" or impactor; in fact they ignored tidal forces, though they did comment that while capture was possible (presumably through chaotic perturbations) it would violate the known lack of a substantial lunar iron core, The omission of tidal effects made a certain amount of sense in the BG approach because, obviously, we never had an object of Mars' mass in Earth orbit, and the intent was to provide initial conditions for the GI theory. Chaotic effects were found to be sufficient to induce collision. Nevertheless, we question here how tidal effects, which were shown by Gerstenkorn (1969) hereinafter G69, by O72, and by Conway (1982) to suffice for actual capture of the Moon, an object obviously having 1 lunar mass, could safely be ignored for an approaching orbiter 8 times more massive.

Tidal forces can be divided into three kinds: a radial force $F_{rPM}$ due to deformation of and dissipation in the PM, a radial force $F_{rE}$ due to deformation of the Earth, and a tangential force $F_{tE}$ due to deformation of the Earth (the ordinary "tidal force" now driving the Moon outwards.) The latter two forces are proportional to the square of the orbiter's mass, so that *per unit orbiter mass* they are *linearly proportional* to mass. The radial tidal force due to dissipation in the satellite is only linearly proportional to its mass (G69) and tends to dominate the tangential tidal force for a PM mass equal to the Moon's (today). The capture cross section for our more massive PM is expected to exceed Gerstenkorn's. The capture of a Mars mass impactor such as the BG object would be considerably enhanced, capture probability exceeding collisional probability, as the increased force $F_{tE}$ which transfers angular momentum to Earth is enhanced by the square of the Mars-like object's mass, while the *collision* cross section goes up only as its radius. The real problem with capture, however, is the angular momentum. The minimum angular momentum for a contact circular orbit would be $\sim 4.3 \times 10^{35}$ Js, as compared to $3.4\text{-}3.5 \times 10^{35}$ Js actual (3.4 present or $\sim 3.5$ primordial). Of course an orbit with the two objects in contact is absurd; they would be highly deformed and might merge into a flattened system, but angular momentum increases with the square root of the separation of centers so larger capture orbits would have a worse problem. This discussion highlights that the small impact parameters needed for a proto-Moon and Earth collision as derived by BG are small subset out of possible impact parameters most of which would violate angular momentum constraints.

To support these assertions let us compare the radial and tangential tidal forces for the four masses: $M_m \sim M_\oplus / 81$, $M_m \sim M_\oplus / 48$, $M_{PM} \sim M_\oplus / 39$ and $M_{whack} \sim M_\oplus / 10$. In the smallest mass case (present Moon mass) we have only verified that we duplicate the results of Gerstenkorn (G69) using either his equations or Conway's. (in the case of Gerstenkorn it is necessary to interpret the equation for $u_0$ at the top of his page 199 as $u_0 = (1+e)/2e$, supplying the missing parentheses). In the two other cases we are interested to compare the collision cross section to the capture cross-section, at the same time including torque-producing tidal forces. The radius of the Earth is taken as 6371 km, the present mean radius. Since the Earth may still accrete, this might be too large, but since the gravitational settling of iron may not have been complete and the rotation is faster, there are compensating effects. For the Moon we use its present radius 1738 km while for our favorite, the intermediate mass PM, we have produced a model (discussed below) whose radius turns out to be 2063 km. The numbers quoted here, sometimes to four significant figures are self-consistent, but there is a modicum of uncertainty or leeway in an underlying number like the mass of the PM. We considered values from $M_{PM} \sim M_\oplus / 48$ to the Mars mass of BG. The latter has additional problems of quite likely being tidally captured, which is not the case. BG ignore capture mostly based on our not having such a massive object in orbit, but the angular momentum would also be way out of bounds. In any case, for the smallest mass considered, $M_\oplus / 54$, all the rock mantle would have to be driven outwards as it is stripped from the PM. This is difficult to arrange, as fragments torn off the inner edge are rather likely to be driven down to the Earth, though, due to the rapid decrease of the PM's orbital radius, they may re-encounter the PM. If half the original rock mantle were to accrete to Earth, the original PM mass would have been $M_\oplus/14$, quite large, causing angular momentum problems. With our preferred mass, about 20% of the original rock mantle is forced down to Earth, while 80% arrives quite rapidly at ~ 3.8 $R_E$. We have assumed as well that 30% of the angular momentum of the iron ball (PM core) just as it begins to disintegrate, $3.599 \times 10^{33}$ Js has gone out to the surviving rock disk which will become the Moon, the rest being donated to Earth. The original angular momentum of the PM at the start of stripping was $1.2665 \times 10^{34}$ Js.

## Parameterizing the orbits

Originally, the problem was solved in FORTRAN at the St. Mary's University computing centre. After the author's appointment expired, the equations for this problem were then re-solved using *Mathematica*, first in an inertial rectangular system with perturbations estimated on hyperbolic or parabolic geocentric orbits, and then in the same coordinate system as BG, with their equations supplemented by tidal forces so as to effect capture. Tidal forces are significant only inside the Hill radius. Finally, *Mathematica* was used for the stripping and re-condensation calculations, in an inertial geocentric system representing a region from just inside the co-rotation radius down to Earth centroid. These analyses having continued for many years, the *Mathematica* versions have changed from version 4 to version 8. In Table III Hard-Wired values are for any mass ratio, but the last column is specialized to 1:39 for the ratio of proto-lunar to Earth mass.

Table III
Central Definitions for Mass Ratio 1:39

| Name | Definition | Hard-Wired Value | Derived Value |
|---|---|---|---|
| $M_E$ | Earth Mass now | $5.972 \times 10^{24}$ kg | |
| $M_{E0}$ | Original Earth Mass | | $5.8943 \times 10^{24}$ kg |
| $M_M$ | Lunar mass now | $7.348 \times 10^{22}$ kg | |
| $M_{M0}$ | Mass of the PM | | $1.51137 \times 10^{23}$ kg |
| $R_E$ | Earth radius | 6371000 m | |
| $a$ | semi-major axis of PM orbit | | 2.23 - 2.7277 $R_E$ |
| $M_{Hull}$ | mass of the PM during stripping | | varies |
| $M_{MI}$ | present lunar core mass | ~ $1.293 \times 10^{21}$ kg | |
| $f_{rock}$ | fraction of PM mantle -> Earth | | 0.287196 |
| $fl$ | present Fe-Ni fraction of Earth | 0.32184 | |
| $fe$ | Fe-Ni fraction Earth+Moon now | | 0.317931 |
| $R_{MC}$ | present core radius of Moon | ~ 350 km | |
| $R_{MC0}$ | core radius of PM | | $1.16796 \times 10^6$ m |
| $M_{MC0}$ | core mass of the PM | | $4.80512 \times 10^{22}$ kg |
| $\rho_{MC0}$ | assumed PM core density | 7200 kg st$^{-1}$ | |
| $M_{lunarRock0}$ | original proto-lunar rock mass | | $1.03086 \times 10^{23}$ kg |

We now sketch the calculations. Quantities that were not hard-wired are generally functions of the variable $f_{rock}$, but we suppress that dependence in the equations below. The concept is that the more original PM rock (mantle) we deposit on Earth, the more massive the PM needed to be. A very small fraction $f_{rock}$ leads to a perhaps problematical low-mass PM, while too large a value leads us almost to the case of Theia. In the low-mass case, the PM mantle tends to re-assemble into the Moon just outside the co-rotation radius, which makes the calculation too sensitive to that radius and to the assumed initial Earth spin rate. It is also unreasonable that all of the PM mantle was driven outside the Roche limit, with none deposited on Earth. A large-mass PM tends to load the early Earth with too much "late veneer." both rock and iron alloy, because the PM has to have a cosmically consistent iron fraction $fe$. I omit a few obvious equations such as that the PM core mass is volume times density.

The basic equations are:

Mass Budget

$$M_E + M_M = M_{E0} + M_{M0} \tag{2.1}$$

The value of $M_{MI}$ does not enter the calculations, other than to allow the original rock density to have been a little larger than the final. The present core is secondary.

Rock fractions

$$M_E(1 - f1) = M_{E0}(1 - fe) \tag{2.2}$$
$$M_{lunarRock0} = (1 - fe) M_{M0} \tag{2.3}$$

Donation of rock to Earth
$$M_E = M_{E0} + f_{rock} \times M_{lunarRock0} \tag{2.4}$$

Iron fractions

$$f1 \times M_E = fe(M_{E0} + M_{M0}) \tag{2.5}$$

Given these equations, the rock fraction donated to Earth, $f_{rock}$ can be adjusted to provide a ratio of the Earth mass to the PM mass over a wide range. Choosing $M_{E0}/M_{M0} = 39$, we find that $f_{rock} = 0.2872$. Fig. (8) shows the dependence. For the case that all the rock mantle of the PM goes into the "final product" we would get a mass ratio ~54. The scenario is very hard to arrange, because it is assumed in that case that *all* the stripped rock moves outward, away from Earth, but we know that some was stripped off the lobe of the PM nearest Earth, and, though the PM Hull moves in so that it might re-encounter those pieces, it can hardly be a 100% re-capture. The planet-ring interaction equations of Goldreich and Tremaine (1980) in fact indicate that a ring of material shed from the Earthward side of the PM departs from it at many km s$^{-1}$ velocity. Furthermore, for mass ratio ~54 the stripped rock ends up just a hair outside the co-rotation radius, a perilous situation, because for small changes in the parameters (remember - we set the inclination of the PM's orbit to 0), we might have the rock that was supposed to form the Moon instead descending to Earth.

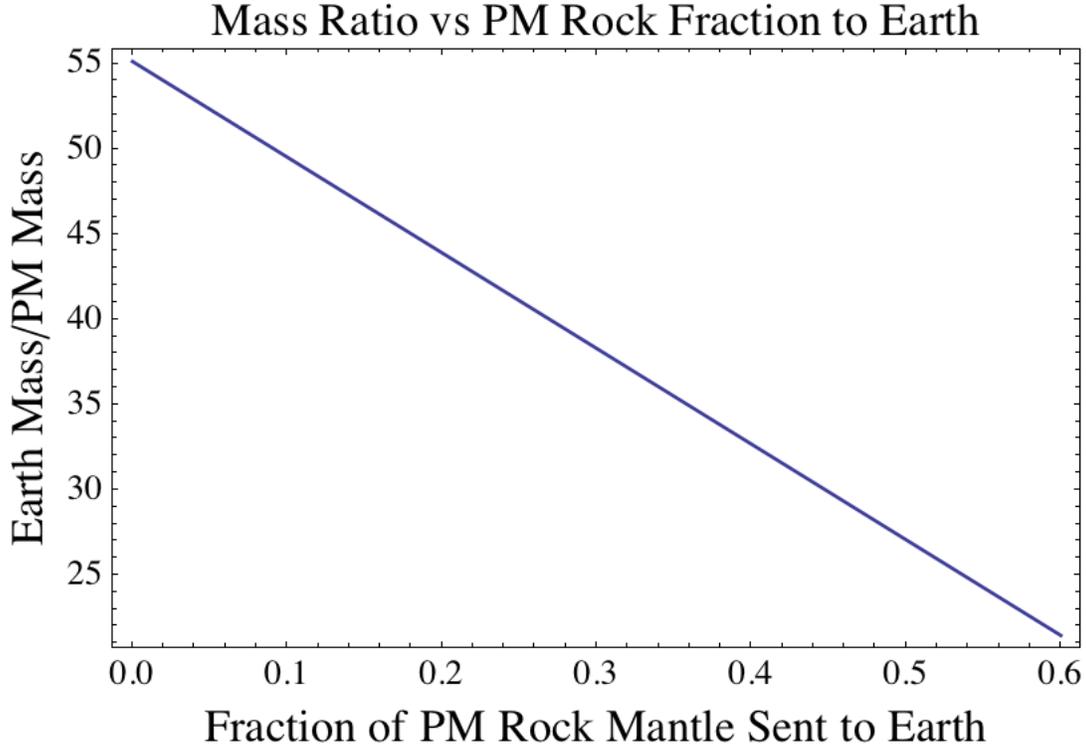

**Figure 9**. The ratio of Earth to PM mass required to produce the actual Moon from the PM. Solutions in a range near 39:1 are viable, assuming no mass is lost, as is the case in this paper.

If we pass to the right edge of Fig. (9), where 60% of the PM rock mantle hits Earth, we get a very large Late Veneer. It is not just the increase in $f_{rock}$ itself that causes this problem, but the whole PM mass has to be larger as well, so we get more rock *and* iron in the Late Veneer. The reason that the iron mass increases when $f_{rock}$ increases is that we assume the cosmic ratio of 32% iron. We feel that the intermediate case of mass ratio 39 is a good compromise. The solution of Eqs. (2.1) - (2.5) for the primordial Earth mass is

$$M_{E0} = \frac{(M_E + M_M)(M_E(1 - f1 - f_{rock}) + f1 M_E f_{rock} - M_M f_{rock})}{(M_E(f1 - 1) - M_M)(f_{rock} - 1)} \tag{2.6}$$

which is plotted below. One also has to derive the mass of the PM from the rock fraction, viz:

$$M_{M0} = \frac{M_M(M_E + M_M)}{(M_E(1 - f1) + M_M)(1 - f_{rock})} \tag{2.7}$$

The solutions are from *Mathematica.* A model of the PM was prepared; to fit the mass ratio 39, with the core and rock densities given, the outer radius before stripping is 0.327866 $R_E$, and the core radius is 0.183324 $R_E$. This may seem large, but please remember that the PM is ~32% iron core! The effects of the compressibility of rock and of iron on the structure of the PM were worked out but the differences from a model of two homogeneous layers proved unimportant to the accuracies we deal with here. The mean density of the PM is 3958.9 kg sr$^{-1}$. for the chosen value $f_{rock}$ = 0.2872, which is keyed to mass ratio $M_{M0} : M_{E0}$ = 1:39.

To summarize, the equations for the capture phase are those of BG, enhanced with the $J_2$ force and tidal forces, described below. The capture phase includes orbit circularization. The stripping phase is handled with tidal torque forces but no radial tidal force, as the orbits are a very tightly wound spiral, about 2000 orbits in ~500 days. These numbers are sensitive to the assumed Earth response time to tidal distortion and could vary quite a bit. The material in the PM Hull is stripped down according to the usual Roche limit equation:

$$RL = cRoche \times R_E \sqrt[3]{\rho_E / \rho_{Hull}} \tag{2.8}$$

where Murray & Dermott (1999) give a value $cRoche$ = 1.44 for a rigid, solid ball and 2.46 for a fluid body filling its Roche lobe. The actual situation is probably closer to the fluid body case, being dynamic, however, as initially the Earth distance changes on the timescale for the PM Hull to deform. We have also not specified the rotation of the Hull about its axis. A compromise constant *cRoche* = 2.44 was adopted, in view of these uncertainties.

## Integrating the orbits

We have performed three kinds of orbital integrations. To critique the Belbruno-Gott theory, we used hyperbolic, but nearly parabolic orbits like theirs and estimated tidal losses for their Mars-mass "Theia" by integrating over the unperturbed orbit. For actual capture of our PM, we used the rotating coordinates of BG, but allowed departure from Keplerian hyperbolae as dictated by Newton's laws, including tidal forces as in Conway (1982) [basically from Mignard (1979, 1980)]. In all cases we have checked that the tidal lag angle is less than 45° (Meyer et al 2011).

A second set of orbits for estimating the tidal torques was done in geocentric inertial coordinates, parameterized by velocity at infinity $V_\infty$, and impact parameter *b*. The latter is the "miss distance" of the asymptote to the original hyperbolic orbit from the Earth's center. G69 found capture of a lunar mass object for $V_\infty \leq 800 \text{ms}^{-1}$, but to jibe with BG we considered velocities in the range 75 – 358 m/s and for generality even ran one case with $V_\infty$ as large at 3 km s$^{-1}$. Although the capture probability is somewhat larger for a

Mars mass PM, BG did not consider capture, and, indeed ignored tidal forces. If the orbit is described by

$$\frac{x^2}{a^2} - \frac{y^2}{b^2} = 1 \qquad (2.9)$$

the impact parameter is $b$. Obviously if $b_{capt}$ is the largest $b$ value for which capture occurs, then the capture probability is $P_{capt} \propto Area = \pi(b_{capt}^2 - b_{coll}^2)$, where $b_{coll}$ is the collision cross-section. Other parameters are the semi-latus rectum $L$ in, the semi-major axis $a = \mu / V_{inf}^2$ where $\mu = G \times (M_{PM} + M_E)$. The perigee distance $r_p$ must exceed the "contact distance" or sum of the two radii, $R_\oplus$ and $R_{PM}$. The eccentricity is given by $\varepsilon = 1 + r_p / a$, and the impact parameter is $b = a\sqrt{\varepsilon^2 - 1}$. The perigee distance is $r_p = a(1-\varepsilon) = L/(1+\varepsilon)$ (I standardize to a > 0, though some authors use the opposite convention for hyperbolic orbits). The collision cross-section is found by setting $r_p = r_{contact} = (R_\oplus + R_{PM})$ which leads to

$$b_{capt}^2 = 2ar_p + r_p^2. \qquad (2.10)$$

This is as far as one can go without calculating the energy losses to tides, which establish $b_{capt}$ in terms of the input parameters such as $V_\infty$ and $b$. To use (2.10), one must set the final energy to zero. Otherwise this equation just duplicates the previous result $b = a\sqrt{\varepsilon^2 - 1}$. In passing to this calculation it is worth noting that the radial tides dominate the tangential (G69) but the former, when taken as *specific* forces (i.e. accelerations) are independent of the satellite mass, while the later depend linearly on it. Thus for more massive PM's such as the one of Martian mass, the rotational tides add in a bit more than for the Moon or our PM. Collisions will occur if $r_p \leq r_{contact}$. Simplistically, one might expect surface disruption in which pieces would be torn off the inner edge of the PM on early passes. This would be a small effect due to dissipative forces in the PM and the short time constant, as shown by Mizuno & Boss (1985). It is just this *difficulty* for the older "dissipative capture" scenarios, as cited by Wood (1986) in his "F" (failure) grade for this alternative, that *enables* our scenario, because it holds the PM together for many passes as the orbit circularizes. Wood thus disposes of disintegrative capture theories with disruption in early near misses, but does not attack intact capture followed by stripping. More recently, Holsapple and Michel (2006) demonstrated that even rubble-pile satellites of zero cohesive strength and density $\rho_{sat}$ can pass within distance $d = 1.5(\rho_{sat} / \rho_P)^{1/3} R_P$ of a planet with density $\rho_P$ and radius $R_P$, or about 61% of the Roche limit. We found that attempting to modify the orbital equations to include changing osculating elements as done by Conway (1982) led to poor modeling of the orbits, so we adopted the BG approach.

**Angular Momentum**

The early Earth rotational velocity is taken as 0.000271331 rad s$^{-1}$ and the equatorial moment of inertia as 9.0 ×10$^{37}$ kg m$^2$. Thus the Earth angular momentum would have been 2.442 ×10$^{34}$ Js. The rotational velocity is not accurate to so many significant figures, but we quote the value used. It is important to have it near that value; too small a value would supply too little angular momentum, while too large a value would put the co-rotation radius (day = month) inside the initial Roche limit. Although the PM approached on a near-parabolic but hyperbolic trajectory, its initial angular momentum would have been fairly small but was enough to boost the total to the Canup estimate 3.5 ×10$^{34}$ Js. The present reference Earth model (Chambat & Valette 2001) has radius of gyration 0.3307 $M_E R_E^2$, where $R_E$ is the mean radius 6371230 m, while the equatorial radius is C = 6378137 m and the polar radius is A = 6356752 m. See also Dziewonski & Anderson (1981). The dynamic form factor $J_2$ is today 0.0010826. We need to model, roughly, how these numbers would have differed at the time of lunar capture. For this, we adopt Eqs. (1) and (2) of Goldreich (1966):

$$I_C = I\left[1 + \frac{2k_s \Omega_\oplus^2 C^5}{9GI}\right] \quad (2.11)$$

$$I_{A,B} = I\left[1 - \frac{k_s \Omega_\oplus^2 C^5}{9GI}\right] \quad (2.12)$$

where the mean moment of inertia (for an equivalent sphere) is $I = M_\oplus R_{gyr}^2$ and $k_s$ is the secular Love number $k_s \approx 0.947$. The secular trend has to include the slow decrease of $I$ due to the radius of gyration factor (0.4 for a uniform sphere) having decreased from perhaps 0.36 to 0.331 as the Earth's core formed. For an initial rotation rate 0.000271331 rad s$^{-1}$, an initial radius of gyration factor 0.33 (where a uniform sphere has factor 0.4) and an Earth angular momentum 2.198 ×10$^{34}$ Js, we obtain an initial $J_2 = 0.0156$. Our initial rotation rate for the Earth leads to a 6.4 hour day before (and, essentially, immediately after) lunar capture in contrast to the 5 hour day of Canup and of Kokubo et al (2000) immediately after their claimed Giant Impact. Their assumed rapid Earth spin puts the co-rotation radius $R_{CO}$ in close at ~2.3 $R_E$, which is desirable for their plan to collect the material that eventually becomes the Moon outside of $R_{CO}$. This leaves the GI theory with a lot of excess angular momentum, which apparently has to go into material lost to the system. Our theory is closed (conservative) in mass. Canup (2004) uses an initial Earth angular momentum 3.5 ×10$^{34}$ Js to match today's Earth-Moon total, ignoring loss to solar torques (Goldreich, Mignard). She is free to make that match because her impactor Theia strikes almost normally, adding little angular momentum. The core and some of the rock from our PM strike the Earth tangentially, adding 1.3 ×10$^{34}$ Js to the angular momentum budget, so our primordial Earth has only 2.2 ×10$^{34}$ Js, leading to an initial day of 6.4 SI hours, as contrasted to Canup's 4 hours.

The transfer of angular momentum (i.e. torque) is shown in Fig. (9). There is a left-to-right mirror-reflection for negative true anomaly, not shown. This figure was generated from an assumed unperturbed orbit, using hyperbolic orbital elements, with velocity at infinity 180 ms$^{-1}$, as in BG. Results for realistic (perturbed) orbits are shown later. In the run generating Fig. (9), no orbital perturbation was allowed; a hyperbolic orbit with near-miss parameters was used. Perigee was at 1.58 $R_E$, clearing collision by 0.263 $R_E$.

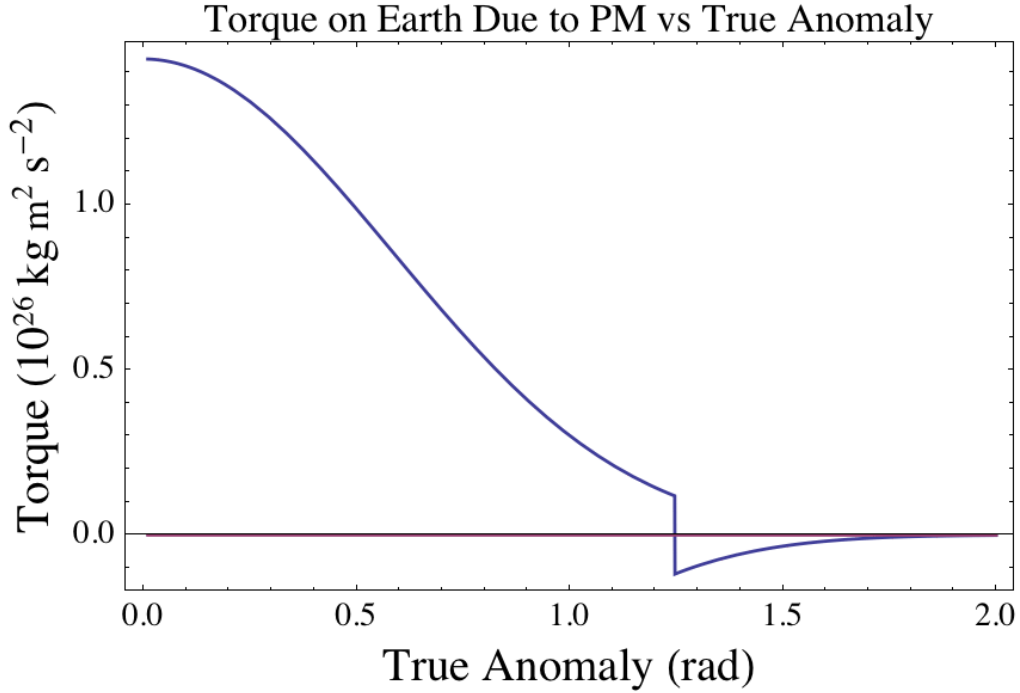

**Figure 9**. Torque by the PM on the Earth, tending to speed its rotation and to foster capture when positive in sign. An idealized orbit was used as if no energy dissipation was present; i.e. Kepler's Laws were applied, ignoring the perturbations.

We have assumed that the PM orbital angular velocity is aligned with the Earth's spin. It can be seen that the effect is to increase Earth rotation, with opposite transfer only outside approximately 1.4 times the radius of perigee. This also implies that perigee will be raised and apogee reduced (as also shown in G69). The angular momentum transfer through the capture period causes only small (1%) changes in Earth rotation rate. For a near pass by Theia, the torques are more severe, about 100 times larger, and are shown in Fig. (10).

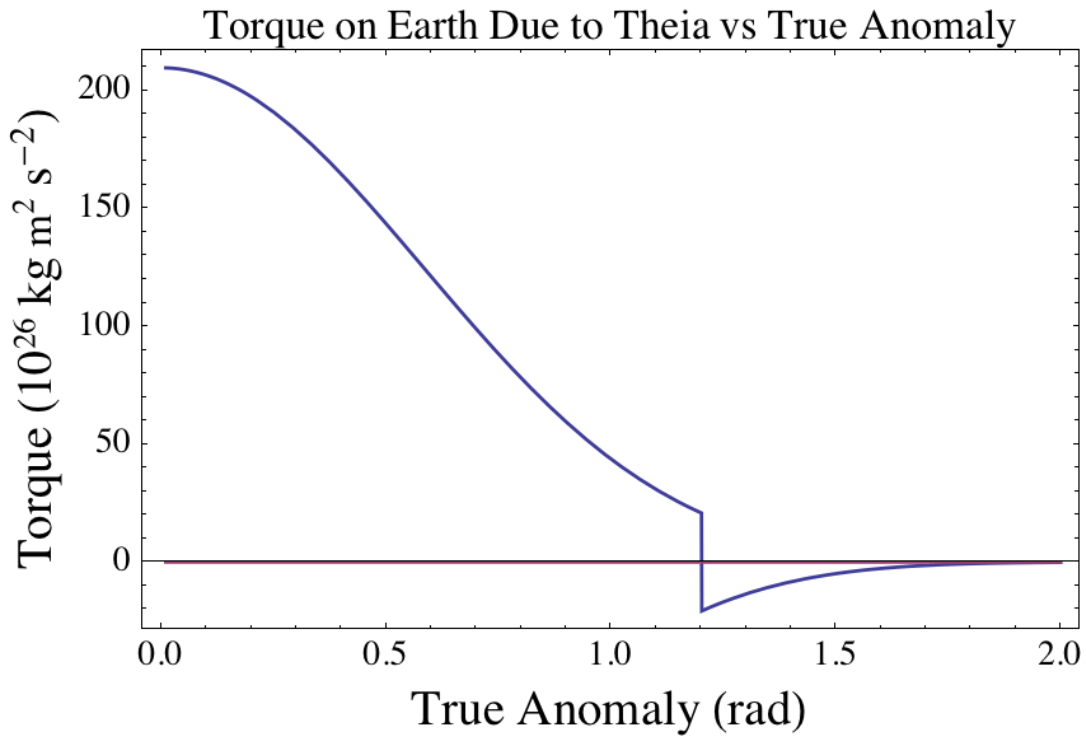

**Figure 10**. Torque by an object with the mass proposed for Theia on the Earth, tending to speed its rotation and to foster capture when positive in sign. An idealized orbit was used as if no energy dissipation was present; i.e. Kepler's Laws were applied, ignoring the perturbations. For this orbit, the perigee is at 1.6857 $R_E$ and the collision radius is 1.532 $R_E$.

## Capture Calculations Starting at L4

The foregoing calculations were all estimates using unperturbed Keplerian orbits to assess the tidal torques. Now we proceed to calculations similar to those in BG, using *Mathematica*, but including tidal forces, which BG omitted. The starting values for the necessary variables were

$x_{start}$ = -0.497664010018950023419785 30794

$vx_{start}$ = 4 × 10$^{-7}$

$y_{start}$ = 0.87468912192389817098594 7457353

$vy_{start}$ = -0.0375931

all in BG units. The variables whose names begin with "v" are velocities, again in BG units. Conservative values $Q \sim 50$ were used for the Earth deformation or tidal lag in early runs, where too large a proto-lunar $k_2 \sim 0.25$ had been used out of analogy with Earth. But the value of $k_2$ value was later set to 0.085 (similar to Mars) both for reasonableness (similarity to Mars) and to avoid excess heating due to radial tidal distortion during orbital eccentricity reduction. Then values $Q \sim 8\text{-}12$ had to be used to effect capture for various initial orbital parameters on leaving L4. These choices of $k2$ and $Q$ may seem unfairly arranged to achieve capture, but in another way one can say we are determining initial parameters of the two bodies, which are only loosely estimated anyway.

We take the radial tidal term from the last term in Conway's Eq. (9), or from G69's Eq.(12) and the tangential from Conway's Eq.(9), terms in $\dot{r}$. The radial specific force terms are

$$force_{1,radial} = -9 k_{2,\oplus} \mu_{PM} R_\oplus^5 \tau_\oplus \dot{r} / r^8 \qquad (2.13)$$

$$force_{2,radial} = -16 k_3 \mu_{PM} R_\oplus^7 \tau_\oplus \dot{r} / r^{10} \qquad (2.14)$$

and

$$force_{3,radial} = -9 k_{2,PM} \mu_\oplus R_{PM}^5 m_\oplus \tau_{PM} \dot{r} / \left(m_{PM} r^8\right) \qquad (2.15)$$

The ratio of the last of these to the first is ~0.5 for the present Moon, ~1.0 for the PM of mass (1/39) the Earth's, and 11.6 for a Mars-like object. The last one itself agrees with G69 if $k_{2,PM} = 1/2$, and we could safely neglect the second, as $k_3$ is small and there are two more powers of the inverse orbital distance $r$ divided by the Earth radius, but we keep it, setting $k_3 = 0.13$. Mignard's (1979) Eq.(5) contains a term identical to our Eq.(2.13) but not those in Eqs.(2.14) or (2.15). The tangential specific force terms (ignored by Gerstenkorn as too small to be of interest, but included by Conway) are, in Conway's case

$$force_{1,tangent} = 3 k_2 \mu_{PM} R_\oplus^5 h_{PM} \tau_\oplus / r^9 \qquad (2.16)$$

and

$$force_{2,tangent} = 6 k_3 \mu_{PM} R_\oplus^7 h_{PM} \tau_\oplus / r^{11} \qquad (2.17)$$

where the relevant terrestrial Love numbers are $k_2$ and $k_3$, and $\tau_\oplus$ is the Earth relaxation time ~90 s. $h_{PM}$ stands for the specific angular momentum $\omega r^2$. The subscript PM refers to the Moon in the case of Gerstenkorn or Conway, to my proto-Moon, like the Moon but with an iron core, in my case, and to the orbiter of mass like Mars' in the case of Belbruno & Gott. In Eqs. (2.13) - (2.17) we have remained close to Conway's notation, dealing in specific force; for actual forces multiply by $M_{PM}$. There is an algebraic sign omitted which is identical to that in Eq. (2.18) below.

We found it more convenient to use the expression for the torque given by Murray and Dermott (1999) although for completeness we included a term in the Love number $k_3$, as per Conway. Furthermore, we have modified in the expressions for the tangent forces a step function in the difference of the lunar angular rate and the Earth rotation rate, which shows up in Fig. (1) as jumps in the torque. Since the *Mathematica* routine "NDSolve" for numerical solution of ordinary differential equations choked on a true step function, we modeled it as

$$\text{step}_{Surrogate}(\omega) = \text{erf}\left[ k_{\text{sharp}} * (\omega - \Omega_E) \right] \tag{2.18}$$

where $k_{\text{sharp}} = 6000$ in BG units, "erf" is the error function, and $\omega$ is the orbital angular velocity of the PM. The value 6000 was chosen to give a sharp quasi-step behavior without generating problems that can arise in solving ordinary differential equations in *Mathematica* when discontinuities are present.

In the calculations for inspiraling and stripping, we obtain the torques from

$$N_1 = \frac{-3k_2 G \text{step}_{surrogate}(\omega) M_{M0}^2 R_E^5}{r_{\oplus,PM}^6 Q} \tag{2.19}$$

and the $k_3$ term

$$N_2 = \frac{-6k_3 G \text{step}_{surrogate}(\omega) M_{M0}^2 R_E^7}{r_{\oplus,PM}^8 Q} \tag{2.20}$$

Of course, when performing calculations in the BG coordinate system, we converted all the SI expressions to their units.

While there may be no single fatal objection to the GI theory, various questions have been raised based on isotopic studies and timescales (Brandon 2007 and references therein, Spicuzza et al 2007, Schmitt 2002, 2006); also, the details of how to make a rocky Moon out of vaporized material from the impactor and Earth mantle may be challenging. Watters et al (2012) present evidence disfavoring a Moon condensed out of vapor. Isotopic inhomogeneities in the Earth's mantle (Kleine 2011), Campbell & O'Neill 2012) argue against complete melting of the Earth as is caused in the GI theory. The GI theory differs from Gerstenkorn's in that the captured object is larger, with an iron core, and was formed at the L4 point, rather than having come from another more obscure provenance. The scenario I propose has the proto-Moon captured in prograde Earth orbit at geocentric radius $a_{capt} \sim 2.702$, with the Earth rotating rather slower than the PM orbited. The proto-Moon's orbital angular velocity would have been 0.00028077 rad sec$^{-1}$, while the initial Earth rotational velocity $\Omega_{E,0}$ would have been 0.000266 to perhaps ~ 0.0003 rad s$^{-1}$, as compared with 0.00007292 now; i.e. the Earth's angular velocity would initially have been about 3.7 times higher. Importantly, then, the PM's orbital angular velocity would have exceeded Earth's at initial perigee and after circularization, but the initial apogee angular velocity just after capture would have been quite small – much less than the Earth's rotational velocity. The exact location of the initial apogee, which was reduced from the infinite apogee at near-parabolic velocity, is quite sensitive to the capture details and is not very important, as the orbit soon circularizes. The sensitivity to velocity on leaving L4 is emphasized by BG, who state "infinitesimally small changes in $V(0)$ at L4 can cause slightly different Earth flyby conditions, which...can cause the trajectories to change noticeably..." We admit to tweaking the initial coordinates and velocity near those at L4 to achieve capture. A random number generator was used (on a small scale in initial position and velocity) to seek the best initial conditions, but tiny changes inserted by hand were also used. It was felt that using the BG orbital equations (in 2D) was essential to constructing realistic orbits, and we produced hundreds of figures like theirs. It is only well inside the Hill sphere that tidal effects become important. The effect of Earth oblateness on the gravitational attraction was modeled rather late in our work, but it turned out to spoil the capture process (which became a near miss, not a collision, for obscure reasons) so the initial conditions at L4 had to be "tweaked" again. We are surprised that for their much shorter terrestrial day ~5 hr, Kokubo et al (2000) neglected the oblateness modification of the gravitational force. Oblateness effects were likewise ignored by BG, who do not appear to have provided an initial spin value or day length. They did include very small, stochastic forces, which we have not. This may relate to our difficulty in finding additional late close encounters after a "near miss" (pass outside the co-rotation radius, or too fast for tidal capture).

The processes are all supposed to be symmetric about the Earth's equatorial plane. This would leave the finally formed Moon in that plane as well, but the "evection" and "eviction" mechanisms may work to correct that during later orbital evolution (Touma & Wisdom 1994, Touma 2000). Rubincam (1975) also discussed a resonance at distance 3.83 $R_E$ where the Moon's inclination could be increased.

How do we get the tidal forces to do an adequate job for the capture into the aforementioned state? Here it is worthwhile to distinguish tangential and radial tidal forces. The commonly considered tangential tidal force $f_{tid,t}$ (Goldreich 1963, Gerstenkorn 1955, Mignard 1979, 1980) works slowly but steadily to transfer angular momentum between a satellite and a planet. In our case, it peaks at perigee in such a way as to reduce apogee, or circularize the orbit. But there is also a *radial* tidal force $f_{tid,r}$ which can be effective at sapping energy (and thus reducing eccentricity) (Mignard 1979, 1980, G69, Conway 1982). The radial force does not, to first order, affect angular momentum, but it damps radial motion, as it is always opposed to the radial velocity. To transfer the PM from a near-parabolic orbit to elliptical requires a velocity "tick" cutting the velocity around perigee, followed by tidal action of both kinds – radial and transverse. Touma (2000) has an expression for the radial force but soon drops that term, though he deals with large changes in eccentricity. He was mostly concerned with inclination, but there is coupling between eccentricity and inclination because torques (such as the solar torque) transfer more angular momentum near lunar apogee than at perigee, while the opposite is true for the torque due to the Earth's $J_2$. The most comprehensive materials on tidal friction are in Efroimsky and Williams (2009). That work is needed for detailed studies where the orbital parameters and internal constitution of the bodies is known accurately, but for the present study the works of Mignard and of Conway suffice.

The capture process can be compared with the BG collision process. The tidal forces per unit impactor or PM mass are proportional to that very mass. Given that Gerstenkorn (1955, 1969) found capture of an object of mass 1/81 that of Earth, namely our Moon, one concludes that the likelihood of capture of a PM of mass ME/39 is greater than Gerstenkorn's and the likelihood of capture of a Mars mass PM is vastly greater. BG base their target radius $b_m$ on a Mars-like PM that suffers random perturbations, and find that, in case of misses, after ~500 y "the flyby distance becomes steadily larger." This agrees with the stochastic nature of perturbations some of which enhance and some of which reduce collision probability. For orbital capture, tidal forces cause monotonic loss of energy and proto-lunar angular momentum, enhancing the process.

## Sensitivity Tests

We have a considerable number of parameters that have to be assigned values to bring the calculations to a result. We admit to "tweaking" parameters within reasonable bounds to get the processes to proceed as we expect. One key need is to have the co-rotation radius outside the initial Roche limit, so that the PM is captured whole and then disintegrates. This may not be absolutely essential because of the demonstration by Holsapple and Michel (2006, 2008) that a brief pass through the Roche zone does not necessarily disrupt a satellite. Nevertheless, in our case, there are repeated excursions across both the co-rotation and Roche boundaries as the orbits circularize, so we prefer the aforementioned nesting. This would be touch-and-go except that our initial Earth rotation is much slower than in the GI theory, and that $J_2$ both leads to a slower rotation for a given angular momentum and also increases slightly the orbital rate of the PM.

The sensitivity to the $Q$ value used to derive the inspiral (along with stripping), which can range from 20 to 100, only changes the time rates for orbital evolution but not the overall dynamics. The equivalent $\tau Earth$ used in the capture process is more critical, because, together with $k2_{PM}$, it determines the relative heating of the Earth and PM. The initial capture process is remarkably insensitive to parameters such as the initial Earth rotation speed and $\tau Earth$ evidently because, after including $J_2$ term in the gravity so as to modify the BG force equation, we re-tuned the initial position (near L4) and velocity to re-establish capture, and in doing so, set up a very close pass, nearing solid Earth (but not colliding with it!) so as to be within any reasonable co-rotation radius.

## Energy and Angular Momentum Budgets

The incident PM (from L4) arrives at speed ~ 180 m/s, thus having energy ~ $1.7 \times 10^{27}$ J, modest compared to the negative total orbital energy when stripping starts. That energy is -$1.2 \times 10^{30}$ J. The orbital energy of the lunar core at 2.23 Earth radii is -$4.7 \times 10^{29}$ J and that of the ring, or newly formed Moon at ~3 Earth radii is -$7.6 \times 10^{29}$ J. From start to finish, the Earth's rotational energy decreases from $3.21 \times 10^{30}$ J to $3.15 \times 10^{29}$ J, which means a loss to tides of $6 \times 10^{28}$ J. This is between 5 and 6 times the loss by the lunar components (core and disc or rings). Nobody gains here, because the lunar mantle and crust have gone only from 2.79 to 3.8 $R_E$, while the core descended steeply to 2.33 $R_E$.

The Earth's initial angular momentum is $2.20 \times 10^{34}$ Js, reducing to $2.14 \times 10^{34}$ Js just when the Moon has newly re-formed. Of course, tidal evolution after that transfers a lot of angular momentum to the Moon, which ends up with more than Earth, as is well known. There can be no overall gain or loss of angular momentum (ignoring Solar tides) and the budget was checked to verify that Earth's loss is Moon's gain. The solar tides are important only for the billions of years since lunar formation. We ignored angular momentum from the spin of the incident PM.

## Roche limit effects

Most of the previous work on the early lunar orbit (Gerstenkorn 1955, 1968, Goldreich 1966, Mignard 1979, 1980, 1981) has traced the orbital evolution backwards from the present epoch, always with the result that the Moon was originally close to the Roche limit. In a pure capture theory, this exercise is useful; for example, Gerstenkorn (1955) found that the Moon arrived on a nearly polar retrograde orbit and swung to an orbit of lower inclination, initiating tidal recession. In the theory presented here, irreversible processes exist – tidal stripping and then re-accretion outside the Roche limit, vitiating attempts to integrate the orbit backwards past ~ 3 $R_E$, where the re-formation of rock crumbs or dust into the Moon started. With considerable foresight – one might say "prophetically," Gerstenkorn (1955, 1969) noted that the Moon would briefly (but repeatedly) enter the Roche limit as calculated by him, at about 2.89 $R_E$, suggesting damage or breakup! Fortunately, Gerstenkorn's Roche limit was based on a low density ~ 3340 kg st$^{-1}$, appropriate for the true Moon as it is today, while our Roche limit for the PM is 2.456 $R_E$, based on the density in the range ~ 4186 to 4221 kg st$^{-1}$, similar to that of an uncompressed the Earth – i.e. the mean density of Earth minerals in the absence of gravity. Thus, in its post-capture orbit, the PM will *not* disintegrate until it is dragged inwards by Earth tides. A bit more elucidation may be helpful in that context: How is it that the Earth tide at perigee will lower apogee, but that at apogee will not initially lower the perigee to inside the Roche limit? The explanation is that at apogee the PM will fall behind the tidal bulge on Earth, so it is driven forward, slightly raising the perigee. An excellent diagram illustrating the switching between forward and reverse torques to the lunar orbit appears as Fig 6 in Singer (1968). Singer on his pp. 220 – 221 actually outlines a theory a little like ours, except that it does not deal with the proto-Moon as a differentiated object, which allows us to strip the outer layers first, these being driven outward by the shepherding action of the remaining mantle and core. Singer deals with tidal perturbations to the radial force but not terms proportional to the radial velocity as do Gerstenkorn, Conway and the author. These terms, being dissipative, are vital to any capture theory.

The Roche problem has been re-evaluated by Holsapple and Michel (2008) and by Walsh & Richardson (2006), taking into account more realistic material properties than Roche's liquid. The net effect is to move the Roche limit inward, which is to our advantage as it is then easier to obtain capture into an orbit inside the synchronous or co-rotation distance before stripping starts.

The disintegrative capture of the Moon has been discussed many times before and it is worth consulting references cited by Boss & Peale (1986) in this regard. All the earlier disintegrative capture theories involved breaking pieces off the Moon on an impossibly short timescale – i.e. a single pass. Cohesive forces and crack propagation delay limit disruption on short timescales. Cohesive forces can also defeat the Roche disruption process for small bodies (Dobrovolskis & Burns 1980, Porco et al, 2007). In the present theory, the PM is expected to be plastic after capture, due to heating during capture. Importantly, we expect the major orbital changes to be caused by the "tangential" tides (at right angles to the radius), while Gerstenkorn, though aware of these, estimated severe heating to > 1000 K. The radial tides dissipate energy mostly in the Moon, while tangential tides dissipate energy in the Earth. Thus, we do not find such extreme heating as Gerstenkorn does during capture. Thermal effects during the post-capture disintegration are under study, but the heating is ameliorated by radiation into space. The latter has little effect during capture, but the surface-to-volume ratio is much bigger for the rock that has been torn off. Anyway, cohesive stresses of order ~ 200 MPa will be much less than pressure ~ 1 to 8 GPa. The existing discussions of disintegrative capture also implied a disorderly breakup, whilst our analysis provides an orderly way to separate rock and iron.

## The Stripping and Shepherding Process

The analysis transitioned to an entirely different family of *Mathematica* workbooks upon capture a little inside the co-rotation radius. The coordinates are rectangular, and approximately inertial in the sense that the Earth-PM center of mass is fixed. Occasionally, to trace the movement of fragments, polar coordinates were used. It proved surprisingly important to account for the angular velocity of the Hull including the motion of Earth *about the common* CM. It was impossible within time constraints to model fragmentation in detail. Some attempts were made, but the fragments proved to be susceptible to slingshot-like ejection from the system due to spurious gravitational interactions with the Hull; in fact those fragments would have re-accreted, but that could not be readily modeled. Instead, we use angular momentum conservation, which allows us to estimate the amount of rock sent in towards Earth, as a fraction of the total PM mantle rock, when we use mass conservation laws and the fact that the PM had an iron core that comprised 32% of its mass.

In our actual *Mathematica* runs the value of the Hull mass as the PM is stripped is primary. Fig. (11) shows the dependence of $M_{Hull}$ on orbit radius (after circularization). The Hull starts out (at the right) with initial mass and loses mass from its mantle until the core is reached at 2.237 $R_E$ orbit radius.

## Motion of the rock disk outward prior to collection of rocks and dust into the Moon

The material stripped from the Hull as shown in Fig. (11) is driven outwards (away from Earth center) by the tidal action of the Hull itself at semi-major axis *a* if it starts outside, or inwards if inside. These motions are calculated ignoring Solar tides and using, again, our Eq. (1.5) from Goldreich and Tremaine (1980). We continue using axial symmetry and the inner and outer disks are modeled as a collection of rings. Six "outside" rings (as in O72) were modeled, and limited modeling of six "inside" rings was done. These are forced down to Earth and in the interim their main effect is to slow the descent of the Hull or iron ball. It is tempting to assign ring masses in the ratio (1-$f_{rock}$):$f_{rock}$ but that leads to a problem with the last-shed outer ring, which does not reach the co-rotation or Roche radii, as it was shed so close to Earth and there is so little time for the Hull or iron ball to drive it out.

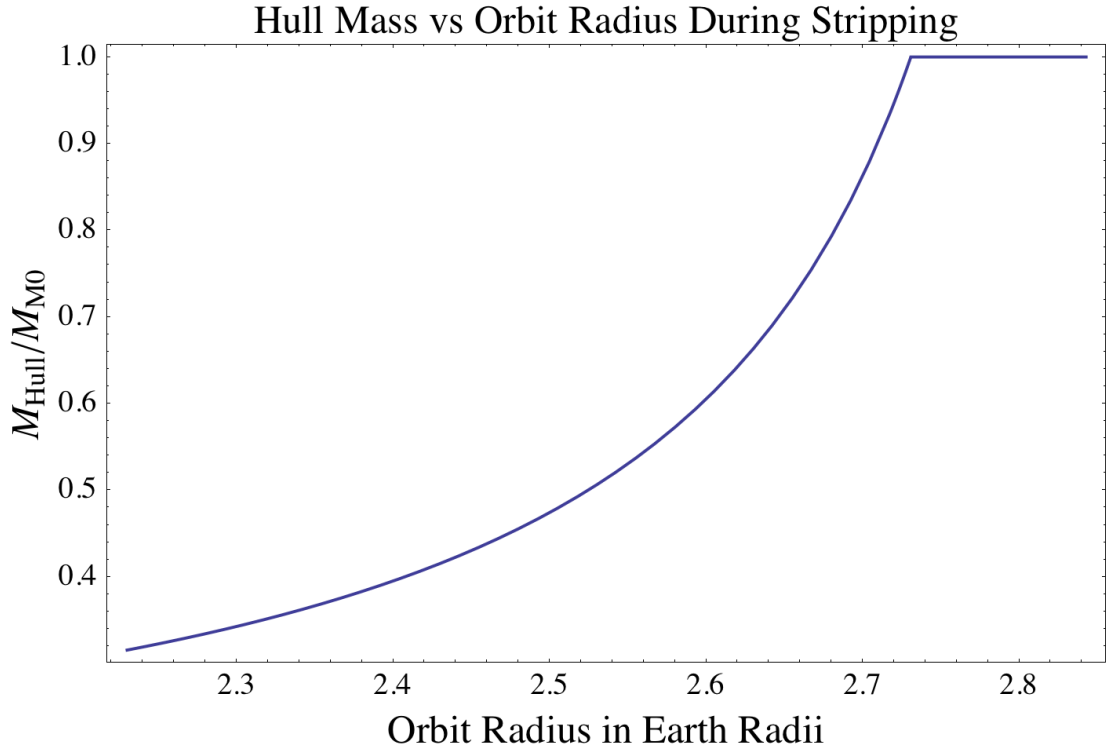

**Figure 11**. Mass of the proto-lunar Hull as rock is gradually stripped. Time increases from right to left. The dependence shown here assumes simple application of Eq.(2.8); see the discussion in connection with Fig. (8)

The destinations of the 6 outer rings using the estimate $\tau_\oplus = 140$ for the relaxation time of the primitive Earth are shown in Fig. (12). The 5th ring makes it to the zone where the Moon will collect, but the 6th fails. Therefore the ratio of masses of outer to inner rings is increased, but without prejudice to the estimate assumed value of $f_{rock}$ since the 6th outer ring will end up at Earth surface.

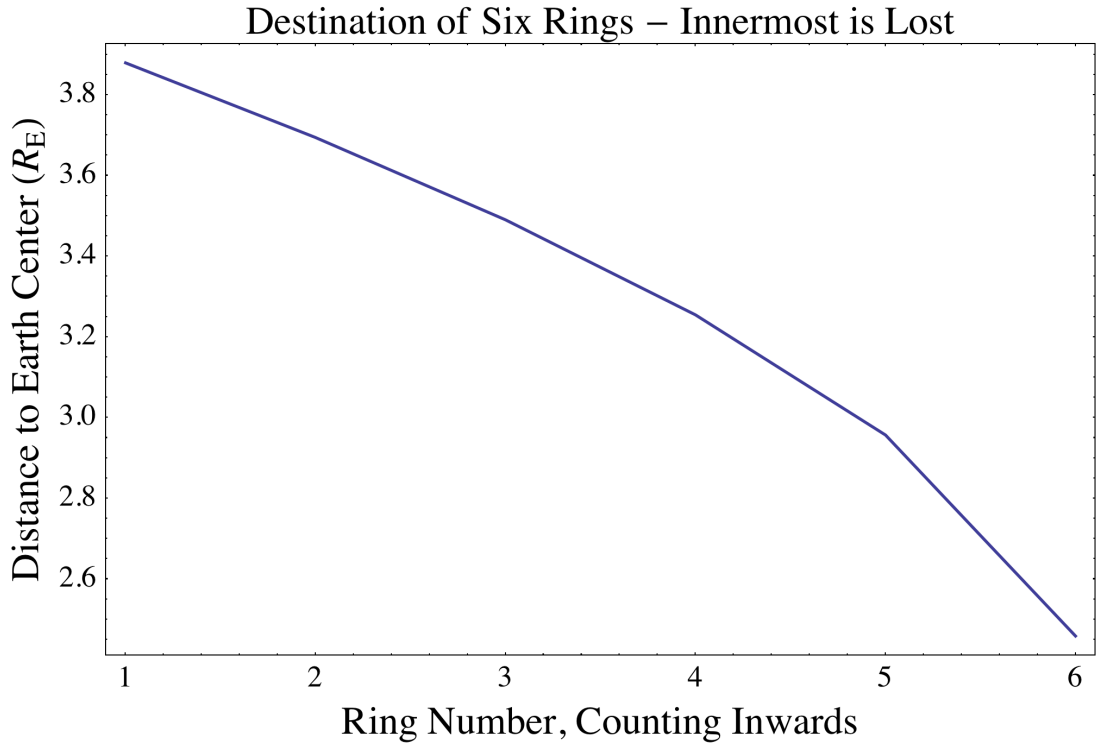

**Figure 12**. End positions of six example rings representing the rock disk external to the orbit of the PM Hull. Five of them reach "safety" outside the Roche limit, where they will re-accrete into the Moon.

The calculations were repeated for double the Earth relaxation time $\tau_\oplus = 280$ - more nearly that of G69. Destinations of the 6 outer rings were again found and are similar to the positions $\tau_\oplus = 140$. The Hull descends faster for the larger value of $\tau_\oplus$ so the time available for the rings to move out is less, and they end up a little closer to Earth. The 5th ring makes it to the zone where the Moon will collect, but the 6th fails. Therefore the ratio of masses of outer to inner rings is increased, but without prejudice to the estimate assumed value of $f_{rock}$ since the 6th outer ring will end up at Earth surface.

These calculations were generally done assuming that the Hull moves in due to Earth torque only, which is not a bad assumption, as the rings move away from the Hull so fast. In most cases, even the motion of the Hull was frozen while a ring moved, but selected cases were repeated with Hull descent included. The final position of the first ring is rather far out, but the model ignores the three-dimensionality of the rock disk. There will be vertical thickness and spreading of the rings, especially the first few, with secant losses diminishing the velocities calculated by the Goldreich-Tremaine equations. The transit times to arrival are comparable to the total stripping time, as shows for the 3rd ring in Fig. (13). The inner material (Hull or ball, inner rock rings) will probably strike Earth in less than the 700 days shown - the time period for inward migration of the Hull and its stripping is quite sensitive to the assumed Earth tidal response time $\tau_\oplus$, which can only be guessed - we have tried values in the range 80 - 500. Because most functions (such as Hull mass) are functions of the orbit radius $a$, the travel times were calculated from the reciprocal of the Goldreich and Tremaine (1980) dependence of radial speed on time; thus, when the radial ring speed is

$$\dot{r} = 0.798 a \sqrt{\frac{r}{a}} \left(\frac{M_{Hull}}{M_\oplus}\right)^2 \left(\frac{a}{a-r}\right)^4 \Omega \tag{2.21}$$

we find the transit time from $dt = dr/\dot{r}$. There is no singularity at $r = a$ because the ring must start one Hull radius from Hull center.

The calculations were repeated for $\tau_\oplus = 140$ and $\tau_\oplus = 280$. The slower Earth response leads to a greater lag angle and thus more torque (this is true up to a lag angle of $45°$ - see Meyer et al 2011). The travel time is now halved, due to the larger torque.

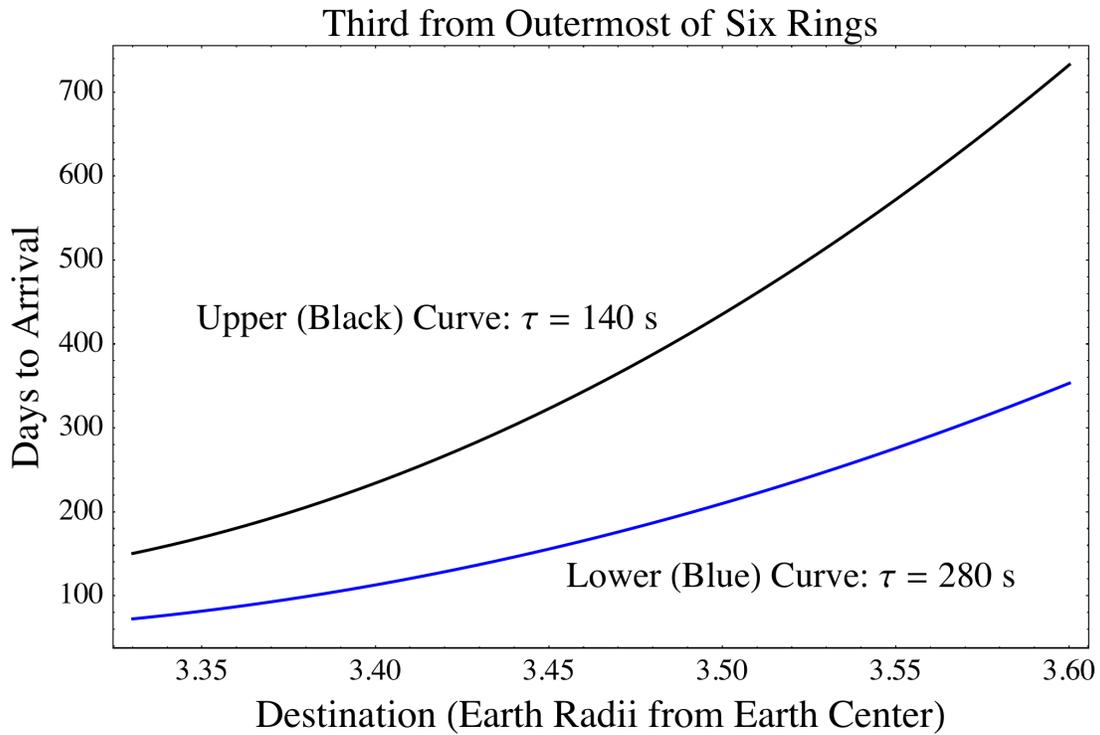

**Figure 13**. Time required for the third of six rings to arrive outside 3.6 Earth radii. The result is sensitive to the Earth response time to the tidal force of the remaining PM (see text). $\tau$ refers to $\tau_\oplus$

## Conclusions

Our Moon was created in a rather complex process (the "GE" event) from a captured Proto-Moon (PM) originally formed at L4 (or L5) at about the same time as the Earth formed. The PM completed core formation early (Yin et al 2002, Edmunson et al 2009), and developed an internal dynamo similar to Mercury's or Earth's. It escaped L4 due to random and/or tidal forces, most likely at one of three times: 4.46 Ga ago, 4.291 Ga ago, or ~3.8-3.9 Ga ago "take your pick." Our theory is not highly sensitive to the choice among these dates, but our ability to explain early magnetization of lunar rock and lava samples works best with younger ages. After orbiting the Sun as in BG, The PM was captured, due to tidal forces, into Earth orbit slightly inside the co-rotation or synchronous radius and spiraled in, shedding its rock mantle as it did so. Most of these fragments were forced out beyond the Roche limit and synchronous radius, following which they formed into the Moon, more or less as we know it. The remaining rock and iron core accreted to Earth at an extremely flat angle, producing the "Late Veneer." Desirably, our theory does not involve whole-Earth melting. The excess iron in the lunar mantle (Day & Walker 2011) might be explained by an iron excess in the lower layers of the PM, which would have been deposited in the Moon's upper layers due to its having been largely turned inside-out. If the latest suggested date (Case III) for the GE event is chosen, the problem that the Moon has reached its present distance with startling alacrity (Kaula 1968) is ameliorated, as it has traveled outward for only ~3.9 Ga, not > 4.2 Ga. This means that less severe assumptions on the shapes and depths of ocean basins need be made. Further, the existence of a relatively warm upper mantle and crust, without an extremely hot core, is improved, due to the decay of U, $^{40}$K, and Th in the interval from creation of the PM to the GE event. Our theory is consistent with an early Moon that was *not* totally molten (Watters et al 2012), while the GI theory almost certainly produces a molten Moon, since it condenses out of vapor. In closing, we urge the reader to return to the point about excess estimates for the age of the Universe, due to Sandage et al, and to question closely if quoted ages for lunar structures may have resulted from inconclusive association of mineral ages in returned samples with actual basins and craters.

## Predictions for GRAIL

GRAIL (Zuber et al 2011) is now in orbit and is taking data. We claim (mostly keyed to questions TBD in Zuber et al 2011, q.v., and bulletized in parallel to them) that:

- The Moon was initially rather cool, not much over 1200 K, and was never in the form of vapor
- The global-scale asymmetry was due to Earth gravity effects during and soon after lunar accretion and not to the late accretion of a companion Moon (Jutzi and Asphaug 2011)
- There may not have been a lunar magma ocean at all (Borg et al, 2011, Gross et al 2012), but if there was, it started internally due to radioactive heating, and surface manifestations would have been due to convective overturn or intrusion of magma ocean lavas into less dense overlying layers
- The lunar core is secondary, and thus is poor in many siderophilic elements. It may be Silicate and Titanium-Rich (Wieczorek and Zuber 2002)
- There is indeed an undifferentiated lower mantle, and it is composed of rocks that originally lay higher - perhaps even in the crust - of the PM (see my section on whether the Moon was turned inside out.) Since we believe that Station 8 Boulder is a survivor of the GE, we assume that the rocks in the unconsolidated lower mantle may be in pieces as large as a meter or more.
- GRAIL will not confirm any basins older than 4.291 Ga (or, probably, 3.8 Ga) as mentioned by Bottke et al (2009), who state (on their p. 6) that basins older than the South Pole-Aitken Basin (SPA) have been conjectured to exist, and who suggest validation by GRAIL

Finally, we have a "prediction" that is for the Earth, not the Moon per se:

The late veneer on Earth originally was concentrated near the (then) equator and was iron-nickel alloy and siderophiles (Pt, Ir,..., probably with some sulfur) mixed with rock of approximately chondritic composition. Given mantle convection and continental drift, it will be difficult to check this prediction, but there may still be some recognizable linearity descended from a great circle origin amongst the oldest zircons, CAIs and greenstone. There would have been about $3 \times 10^{22}$ kg of rock and $4.8 \times 10^{22}$ kg of iron alloy.

## Acknowledgements

I am indebted to Al Harris for numerous conversations. E. David Skulsky participated in defining an initial version of this theory. Communications from James Day, A. Saal, F. Albarède, Harrison Schmitt, S. Mojzsis, and Lindy Elkins-Tanton were helpful. I am indebted to Paul Renne for an update on Ar-Ar age determinations. The *Mathematica* support staff was quite helpful. My interest in the angular momentum of the Earth-Moon system was initiated by the paper of Fahlman and Anand (1969), a copy of which was initially sent to me by Dr. Anand. I am indebted to the Royal Astronomical Society of Canada for permission to reproduce figure 1 of their paper.

## Dedication

This paper is dedicated to the memory of my father, Julius P. Noerdlinger, 1893-1959.